\documentclass[12pt]{article}
\usepackage{amsmath,amssymb,amsfonts}
\usepackage{graphicx}
\usepackage{epsfig}
\usepackage{sectsty}
\usepackage{cite}
\usepackage{epstopdf}

\makeatletter \@addtoreset{equation}{section} \makeatother

\renewcommand{\thefootnote}{\#\arabic{footnote}}
\def\IR{\mathbb{R}}\def\IZ{\mathbb{Z}}

\def\CA{{\cal A}}
\def\CG{{\cal G}}
\def\CH{{\cal H}}\def\CI{{\cal I}}
\def\CL{{\cal L}}\def\CM{{\cal M}}
\def\CN{{\cal N}}\def\CO{{\cal O}}\def\CP{{\cal P}}
\def\CQ{{\cal Q}}\def\CR{{\cal R}}
\def\CU{{\cal U}}\def\CV{{\cal V}}

\def\CZ{{\cal Z}}

\def\a{\alpha}\def\b{\beta}\def\g{\gamma}
\def\d{\delta}\def\e{\epsilon}

\def\s{\sigma}

\def\D{\Delta}
\def\S{\Sigma}

\def\half{\frac{1}{2}}
\def\thalf{{\textstyle \frac{1}{2}}}

\def\goto{\rightarrow}

\def\p{\partial}
\def\tr{{\rm tr}}

\def\da{{\dot{\a}}}

\newcommand{\be}{\begin{eqnarray}}
\newcommand{\ee}{\end{eqnarray}}
\newcommand{\nn}{\nonumber}
\newcommand{\bn}{\begin{enumerate}}
\newcommand{\en}{\end{enumerate}}

\begin{document}

\begin{titlepage}
\vfill
\begin{flushright}
{\tt\normalsize KIAS-P09009}\\
%{\tt\normalsize SU-ITP-08-21}\\
\end{flushright}
\vfill
\begin{center}
{\Large \bf    Nonrelativistic   Superconformal M2-Brane  Theory}

\vfill
  \, Ki-Myeong Lee$^\dagger$, Sangmin Lee$^\ddag$, Sungjay Lee$^\dagger$
 \footnote{\tt e-mail:
  klee@kias.re.kr, sangmin@snu.ac.kr, sjlee@kias.re.kr}

{\footnotesize\it
\vskip 5mm
${}^\dagger$Korea Institute for Advanced Study, Seoul 130-722, Korea \\
${}^\ddag$Department of Physics, Seoul National University, Seoul 151-747, Korea
}
\vfill
\end{center}

\begin{abstract}
\noindent
We investigate  the low energy physics of particles in the symmetric phase of the $\CN=6$ mass-deformed ABJM theory in terms of the superconformal nonrelativistic field theory with 14 supercharges. They describe
a   certain kind of excitations on M2 branes in the background of external four-form flux.
We study the nonrelativistic superconformal algebra and their representations by using the operator-state correspondence with the related  harmonic oscillator Hamiltonian. We find the unitarity
bounds on the scaling dimension and particle number of any local operator,
and comment on subtleties in computing the superconformal Witten index
that counts the chiral operators.
  \end{abstract}

  \vskip 2cm
  \end{titlepage}

\tableofcontents
\renewcommand{\thefootnote}{\#\arabic{footnote}}
\setcounter{footnote}{0}

\section{Introduction  and Concluding Remarks }

Recently there have been much interests in the construction of gravity
dual for the nonrelativistic  conformal theories
~\cite{Son:2008ye,Balasubramanian:2008dm,Herzog:2008wg,Maldacena:2008wh,Adams:2008wt, Hartnoll:2008rs,Donos:2009en}. The nonrelativistic
conformal symmetry, so-called the Schr\"odinger symmetry
\cite{Hagen:1972pd,Niederer:1972zz}\footnote{Nonrelativisic conformal symmetry
broadly refers to scale invariance under $(t, x_i) \goto (\lambda^z t, \lambda x_i)$ for arbitrary constant $z$ (dynamical exponent).
We will focus exclusively on the $z=2$ case which admits
a larger algebra than a generic $z$.
}, has
appeared in many condensed matter systems, and also in some class of
nonrelativistic field theories~\cite{Nishida:2007pj}.

In last year, Aharony, Bergman, Jafferis, and  Maldacena (ABJM) have
proposed a  three-dimensional superconformal Chern-Simons matter theory as
a theory on multiple  M2   branes in an orbifold $\mathbb{C}^4/Z_k$~\cite{Aharony:2008ug}.  This
ABJM model has $\CN=6$ supersymmetries and  $SU(4)_R\times U(1)$ global
symmetries. This theory has a mass-deformation which preserves
supersymmetry but reduces $SU(4)_R$ symmetry to $SU(2)\times SU(2)\times
U(1)$  and breaks conformal symmetry~\cite{Hosomichi:2008jd,Hosomichi:2008jb,Gomis:2008vc}.
While this deformed theory has discrete vacua~\cite{Gomis:2008vc},
its nonrelativistic limit in the symmetric phase
turns out to acquire a superconformal symmetry which is different
from the original mass-deformed ABJM model.

In this work, we investigate in details the  nonrelativistic superconformal ABJM model.
We write down the theory, which is characterized by the mass parameter
$m$, gauge group $U(N)\times U(N)$, and the Chern-Simons level $k$. We
also find  its symmetries and conserved charges, and study the symmetry
  algebra and its representations. This
nonrelativistic theory describes the low energy dynamics of a number of
massive charged particles, not a mixed set of particles and
anti-particles, in the symmetric phase of the relativistic mass-deformed ABJM model.
In addition to the conformal symmetry, the number of supercharges also increases
from the original 12 to 14. These 14 supercharges get split to 10 kinematical supercharges,
2 dynamical supercharges and 2 conformal supercharges.
We study the representations of the nonrelativistic superconformal (super
Schr\"odinger) algebra in the related many-body theory with a harmonic
potential. There are interesting bounds on the scaling dimension
$\D_\CO$  and the particle number $N_\CO$ of a local operator $\CO$ by its
charges. We also comment on the superconformal Witten index which counts the so-called chiral operators.

In  the symmetric phase of the mass-deformed
ABJM model, there is no massless particles in the symmetric phase and so
a pair of particle and antiparticle cannot annihilate to the vacuum and
would remain as a sort of massive meson state with maybe some
bound energy. Thus one has to decide whether one wants to keep only
particles or both particles and anti-particles in the nonrelativistic
limit, which changes the nonrelativistic theory and its supersymmetries.
In our case, the ABJM model has the global $U(1)$ symmetry to start
with, and so we keep only particles of positive $U(1)$ charge in the low energy dynamics.
If we have kept both particles and antiparticles in the nonrelativistivic dynamics,
all 12 supersymmetries would become kinematical and there would be no additional
conformal supercharges.

Similar question would arise in the study of the nonrelativistic limit of the related   $\CN=8$
Bagger-Lambert-Gustavsson (BLG) model
~\cite{Bagger:2006sk,Bagger:2007jr,Bagger:2007vi,Gustavsson:2007vu,Gustavsson:2008dy}.
The BLG model also has the supersymmetry preserving mass deformation which breaks $SO(8)$
R-symmetry to $SO(4)\times SO(4)$ R-symmetry,
which allows the nonrelativistic limit in the symmetric phase~\cite{Gomis:2008cv,Hosomichi:2008qk}.
As there are no $U(1)$ symmetry in the BLG model, we do not have
a predetermined notion of particles and antiparticles as in the ABJM model.
The straightforward nonrelativistic limit of the BLG model, which keeps all kinds of massive particles,
lead to only kinematical supercharges.
Instead, if we keep particles in the sense of the ABJM model,
then we end up with precisely the same symmetry algebra as in the ABJM model.

It is well-known that the theories with the
$d$-dimensional Schr\"odinger symmetry can be obtained
by performing the discrete light-cone quantization (DLCQ) of
theories with the $(d+1)$-dimensional relativistic
conformal symmetry. As pointed out in~\cite{Maldacena:2008wh},
the DLCQ of field theories however raises
many subtle issues, and as a consequence
%it need a very careful
%analysis to get the physically meaningful quantities including
%the resulting effective nonrelativistic
it seems rather difficult %(if not impossible)
to obtain the explicit field theory Lagrangians.
Recent works
\cite{Son:2008ye,Balasubramanian:2008dm,Herzog:2008wg,Maldacena:2008wh,Adams:2008wt,Hartnoll:2008rs,Donos:2009en}
to construct the supergravity solutions
of interest rely on the DLCQ embedding,
so their interpretation in terms of the dual field theories
remains unclear.

Since our work begins with the ABJM model
with a definite proposal for the gravity dual,
%As far as we know, there no known examples of
%physical supersymmetric gravity solutions with the Schr\"odinger symmetry.
our work may
%lead to an interesting suggestions on the possible gravity
%dual of our theory and
help to build the first concrete example of nonrelativistic holography.
The gravity dual of the ABJM model is $AdS_4\times S^7/Z_k$,
and the mass deformation can be induced by
introducing a certain class of four-form field strength to this
geometry~\cite{Bena:2000zb,Bena:2004jw}.
Now we want to keep only particles in the symmetric
phase. How we can manage this in its  gravitational counter part is not
clear at the moment.

In the relativistic conformal field theories, there has been a natural
correspondence between operators and states by considering the theory
on spheres instead of plane.
This can be achieved by the radial quantization of the Euclidean
time theory as space and time has the same scaling dimension. This leads
to a simple representation of the conformal symmetry algebras.
Since the nonrelativistic conformal theories has anisotropic
scaling behaviors of time and space $(t, x_i) \rightarrow (\lambda^2t,
\lambda x_i)$, one can not simply apply
the above idea in our nonrelativistic ABJM model. However,
the recent work by Nishida and Son~\cite{Nishida:2007pj} has shown that
one can define a new Hamiltonian with a harmonic potential
for a given theory with Schr\"odinger symmetry and find the energy
eigenvalues and states of the system with the harmonic potential
for a given conformal primary operators. We generalize this
scheme of the operator-state correspondence to our supersymmetric
case and get some useful unitarity bound on  scaling dimension and
particle number of any given operator in terms of its other quantum numbers.
The so-called chiral operator saturates the bound of the scaling
dimension.

The key challenge  of our nonrelativistic superconformal field theory
with nonabelian gauge group $U(N)\times U(N)$ would be how to impose the local
Gauss law and how  to find the gauge-invariant operators.
 From the nature of the Gauss law, one can see the elementary physical
fields should carry both charge and magnetic flux of the abelian
$U(1)\times U(1)$. In addition, one should  impose the
non-abelian Gauss law to get a charge-flux composite operators. While
they would be  invariant
under the local nonabelian gauge transformations, we expect that they
would form also nontrivial representations of the global part of $U(N)\times
U(N)$ gauge symmetry. As   charge-flux composite operators, the physical
operators would be also quasi-local. The correlation functions of these
operators and their operator products would be of much physical interest.
It would be interesting to
calculate their  scaling dimensions and correlation functions perturbatively in the
weak coupling limit.

We argue that the resulting charge-flux operators as the gauge invariant creation and annihilation operators for each massive particles are chiral operators at least in the weak coupling limit. Maybe
only certain kind of operator products of these flux-charge composite operators would remain chiral.
One can define the Witten index to count the chiral operators, and these would contribute
to the counting.  This contradicts with the definition of the chiral operators in the
relativistic theory where they should be gauge singlet or invariant. Any physical operators or states of the nonrelativistic theory also needs to be invariant under the local gauge transformation, but this does not mean that, as the flux-charge composite objects, they have to be  singlet under the global part of the gauge transformation.

There has been many studies of the massive Chern-Simons-matter theories
of less supersymmetries and their nonrelativistic counter parts~\cite{Jackiw:1990mb},
and further of the superconformal theory~\cite{Leblanc:1992wu}.
Also a further investigation of super-Schr\"odinger algebra has been also studied in Ref.~\cite{Sakaguchi:2008rx,Sakaguchi:2008ku,Sakaguchi:2008zz}
More recently there has been a series of work by
Nakayama et al on the subject~\cite{Nakayama:2008qm,Nakayama:2008qz}.
This study has led to two classes of nonrelativistic supercharges:
kinematical ones and dynamical ones. Moreover these nonrelativistic
theories has the BPS soliton spectrum~\cite{Dunne:1990qe},
defined by the covariantly holomorphic matter fields satisfying the
Gauss law. In our nonrelativistic model, one see that similar
solitons, if exist, would be classical versions of chiral or anti-chiral operators.

The contents of the paper are organized as follows. In Sec.~2,
we take the nonrelativistic limit of the ABJM model. In Sec.~3, we find many symmetries and the corresponding charges, including
supersymmetries and nonrelativistic superconformal symmetries.
In Sec.~4, we explore the representation of the superalgebra and study
chiral primary operators with some unitarity bounds on scaling dimensions.

\noindent {\bf Note added: } While we were preparing this paper, a preprint
appeared on the arXiv which contains some overlap with our work~\cite{Nakayama:2009cz}.

%%%%%%%%%%%%%%%%%%%%%%%%%%%%%%%%%%%%%%%%%%%%%%
%%%%%%%%%%%%%  NR limit of ABJM  %%%%%%%%%%%%%
%%%%%%%%%%%%%%%%%%%%%%%%%%%%%%%%%%%%%%%%%%%%%%

\section{Nonrelativistic Limit of the  Mass-deformed ABJM Model}

\subsection{The mass-deformed ABJM model}

Let us start with  a short description on the ABJM model
\cite{Aharony:2008ug}, believed to describe the dynamics of
multiple M2-branes probing a certain orbifold geometry. This
${\cal N}=6$ supersymmetric model has the gauge symmetry $G=
U(N)_k \times U(N)_{-k}$ whose gauge fields are denoted by
$A_\mu$ and $\tilde A_\mu$ with the Chern-Simons kinetic term
of level $(k,-k)$.
The bi-fundamental matter fields are composed of four complex scalars
$Z_\a$ ($\a=1,2,3,4$) and four three-dimensional spinors
$\Psi^\a$, both of which transform under the gauge symmetry
as $({\bf N}, \bar{\bf N})$. As well as the gauge symmetry,
the present model also has additional global  $SU(4)_R\times U(1)$
symmetry, under which the scalars $Z_\alpha$ furnish the representation
${\bf 4}_+$ while the fermions $\Psi^\alpha$ furnish ${\bf \bar 4}_+$.

For the clarity we hereafter reintroduce the Planck constant $\hbar$ and
speed of light $c$ in our discussions. The ABJM Lagrangian is made of
several parts
\begin{eqnarray}
  {\cal L}={\cal L}_\text{CS} + {\cal L}_\text{kin} +
  {\cal L}_\text{Yukawa} + {\cal L}_\text{potential}\ ,
\end{eqnarray}
the Chern-Simons and kinetic terms
\begin{eqnarray}
  {\cal L}_\text{CS} + {\cal L}_\text{kin}
  &=& \frac{ k  \hbar c}{4\pi }\e^{\mu\nu\rho}
  \text{Tr} \Big( A_\mu \partial_\nu A_\rho -  \frac{2i}{3}
  A_\mu A_\nu A_\rho  - \tilde{A}_\mu \partial_\nu \tilde{A}_\rho
  +  \frac{2i}{3}  \tilde{A}_\mu \tilde{A}_\nu \tilde{A}_\rho \Big)
  \nonumber \\ &&
  \hspace{0.3cm}- \text{Tr}  \left( D_\mu \bar{Z}^\a D^\mu Z_\a +
  i \bar{\Psi}_\a \g^\mu D_\mu \Psi^\a \right)\  ,
\end{eqnarray}
the Yukawa-like interactions
\begin{eqnarray}
  {\cal L}_\text{Yukawa}  &=&
  \frac{2\pi i}{k\hbar c }\text{Tr}
  \Big(  \bar{Z}^\a Z_\a \bar{\Psi}_\b \Psi^\b
  - Z_\a \bar{Z}^\a \Psi^\b \bar{\Psi}_\b
  +2 Z_\a \bar{Z}^\b \Psi^\a \bar{\Psi}_\b
  -2 \bar Z^\a Z_\b \bar{\Psi}_\a \Psi^\b \nonumber \\
  && \hspace{1.3cm}
  + \epsilon_{\a\b\g\d} \bar{Z}^\a \Psi^\b \bar{Z}^\g \Psi^\d
  -  \epsilon^{\a\b\g\d} Z_\a \bar{\Psi}_\b Z_\g \bar{\Psi}_\d
  \Big)\ ,
\end{eqnarray}
and the sextic scalar interactions
\begin{eqnarray}
   {\cal L}_\text{potential} &=& -
   \frac{4\pi^2}{3k^2 \hbar^2 c^2} \text{Tr}\Big(
   6 Z_\a \bar Z^\a Z_\b \bar{Z}^\g Z_\g \bar Z^\g
   - 4Z_\alpha \bar{Z}^\b Z_\g  \bar Z^\a Z_\b \bar Z^\g \nn \\
   && \hspace{1.9cm}
   - Z_\a \bar Z^\a Z_\b \bar Z^\b Z_\g \bar Z^\g
   - Z_\a\bar Z^\b Z_\b \bar Z^\g Z_\g \bar Z_a  \Big)\ .
\end{eqnarray}
The positive definite potential $\CV_0$ can be expressed in terms of
third order polynomials $W$ and their hermitian conjugates $\bar W$:
\begin{eqnarray}
  \CV= \frac23 \text{Tr}\Big(W^\a_{\b\g} \bar{W}^{\b\g}_\a\Big)
\end{eqnarray}
with
\begin{eqnarray}
  && W_{\b\g}^\a =-  \frac{\pi}{k \hbar c}
  \Big( 2Z_\b \bar Z^\a  Z_\g +\d^\a_\b
  (Z_\g \bar Z^\rho Z_\rho -   Z_\rho\bar Z^\rho Z_\g)
  \Big) - (\b \ \leftrightarrow \  \g)\ , \nn \\
  && \bar W^{\b\g}_\a = + \frac{\pi}{k\hbar c}
  \Big(  2\bar Z^\b Z_\a  \bar Z^\g
  +\d_\a^\b (\bar Z^\g Z_\rho \bar Z^\rho-  \bar Z^\rho  Z_\rho \bar Z^\g)
  \Big)  - (\b \ \leftrightarrow \  \g)\ .
\end{eqnarray}

We basically use the convention of \cite{Hosomichi:2008jb}
except the hermitian gauge fields so that the covariant
derivatives now become
\begin{eqnarray}
  D_\mu Z_\a = \partial_\mu Z_\a - i A_\mu Z_\a + i Z_\a \tilde A_\mu ,
\end{eqnarray}
and the explicit c dependence in temporal Lorentz indices, for examples,
$\partial_0=\partial_t/c$. It is known that this theory is invariant under
the $\mathbb{Z}_2$ symmetry, or parity $Z_\a,\Psi^\a, A_\mu, \tilde{A}_\mu
\leftrightarrow \bar{Z}^\a , \bar{\Psi}_\a, \tilde{A}_\mu, A_\mu $. For
the gauge invariance, the Chern-Simons level $k$ is required to
be integer quantized, and is chosen to be positive $k > 0$ in this paper.
The trace is  over $N\times N$ matrices of
either gauge group and leaves the gauge invariant quantities. The
spinor contraction is the standard one.

This Lagrangian is invariant under the
${\cal N}=6$ supersymmetry whose transformation rules are
\begin{eqnarray}
  \d Z_\a  &=& i \xi_{\a \b} \Psi^\b,  \nonumber \\
  \d \bar{Z}^\a &=& i\xi^{\a\b}\bar{\Psi}_\b, \nonumber  \\
  \d \Psi^\a  &=& -  \g^\mu \xi^{\a\b}  D_\mu Z_\b +W^\a_{\b\g} \xi^{\b\g} ,
  \nonumber \\
  \d \bar{\Psi}_\a &=&  -  \g^\mu  \xi_{\a\b}   D_\mu  \bar{Z}^\b
  + \bar{W}^{\b\g}_\a \xi_{\b\g} , \nn \\
  \d A_\mu &=& + \frac{2\pi }{k \hbar c}
  ( Z_\a\bar{\Psi}_\b \g_\mu \xi^{\a\b} + \Psi^\a \bar Z^\b  \g_\mu\xi_{\a\b}  )
  ,  \nn  \\
  \d \tilde{A}_\mu &=&
  - \frac{2\pi  }{k \hbar c} (\bar\Psi_\a Z_\b \g_\mu \xi^{\a\b}
  +\bar Z^\a\Psi^\b \g_\mu \xi_{\a\b})\  .
  \label{susytrans}
\end{eqnarray}
Here the supersymmetry transformation parameters
$\xi^{\a\b}=-\xi^{\b\a}$ satisfy the relations
\begin{eqnarray}
  \xi_{\a\b} =(\xi^{\alpha\beta})^* = \half \e_{\a\b\g\d} \xi^{\g\d}\ ,
  \label{12susy}
\end{eqnarray}
with $\epsilon_{1234}=\epsilon^{1234}=1$.

It is well-known that the M2-brane theory allows a mass deformation
\cite{Hosomichi:2008jd,Hosomichi:2008jb,Gomis:2008vc}
which preserves whole Poincar\'e supersymmetry.
The mass contribution to the Lagrangian is
\begin{eqnarray}
  \hspace*{-0.4cm}
  \CL_\text{m} =  - \text{Tr}\Big(
  M^\a_{ \ \g} M^\g_{\b}  \bar{Z}^\b Z_\a + i \bar{\Psi}_\a M^\a_{\ \b} \Psi^\b
  \Big) - \frac{4\pi}{k \hbar c}
  \text{Tr} \Big( Z_a \bar{Z}^\a Z_\b \bar{Z}^\g M^\b_{\ \g}
  - \bar{Z}^\a Z_\a\bar{Z}^\b Z_\g M^\g_{\ \b} \Big) ,
\label{massL}
\end{eqnarray}
where the matrix $M^\a_{\ \b}$ satisfies
\begin{eqnarray}
  M^\dagger = M, \qquad \text{Tr}M =0, \qquad M^2 = \left(\frac{mc}{\hbar}
  \right)^2 {\bf 1}_2\ .
\end{eqnarray}
Up to $SU(4)_R$ rotation, one can choose $M$ to be diagonal
\begin{eqnarray}
  M = \frac{mc}{\hbar}~\text{diag} \big( 1,1,-1,-1 \big)\ ,
\end{eqnarray}
which implies that deformation breaks the R-symmetry group down to
$SU(2)\times SU(2)\times U(1)$. The mass-deformed ABJM can be understood
as the deformation of W tensor as
\begin{eqnarray}
  \delta_\text{m} W^\a_{\b\g}
  = \frac{1}{2}\big( M^\a_{\ \b}Z_\g - M^\a_{\ \g} Z_\b \big)\ , \qquad
  \delta_m  \bar{W}^{\b\g}_\a =
  \frac{1}{2} \big(M^\b_{\ \a}\bar Z^\g - M^\g_{\ \a} \bar{Z}^\b \big).
\end{eqnarray}
The total scalar potential, sum of $\CV_0$ and the bosonic part of $\CL_\text{m}$,
can be nicely expressed as complete squares again
\begin{eqnarray}
  \CV_\text{total} = \frac23 \text{Tr}
  \Big[ (W+ \delta_m W)^\a_{\b\g} (\bar{W}  +\d_m \bar{W})^{\b\g}_\a \Big]\ .
  \label{mpot}
\end{eqnarray}
This mass-deformed Lagrangian still preserves $\CN=6$ supersymmetry,
once the fermionic transformation rules (\ref{susytrans}) are
modified as
\begin{eqnarray}
  \delta_m \Psi^\a =  \delta_m W^\a_{\b\g}\xi^{\b\g} , \
  \delta_m  \bar{\Psi}_\a = \delta_m W^{\b\g}_\a  \xi_{\b\g}\ .
\end{eqnarray}

\subsection{A nonrelativistic limit }

The mass-deformed potential for the theory with gauge group $U(N)\times U(N)$
has a discrete set of vacua where the scalar fields takes nonzero expectation
values with different symmetry breaking patterns~\cite{Gomis:2008vc}. Here we are interested
in the symmetric vacuum where the scalar expectation values vanish
and there is no broken gauge or global symmetries. In the symmetric vacuum,
there are only massive charged particles and antiparticles.
One may not expect particle-antiparticle annihilations to massless particles,
even though it is not clear at this moment whether they form any stable bound states.
For a given number of particles and antiparticles, we can consider
the low energy physics where the speed of particles are much slower
than that of the speed of light. There would be no particle or
antiparticle creations from the collisions as the particle momenta
are much smaller than their mass.

There can be many possible nonrelativistic systems obtained from the ABJM model as
one can choose what kinds of particles we want to keep in the nonrelativistic theory.
Thus, the remaining symmetry including the supersymmetry after
the nonrelativistic limit will depends on our choice of particles.
\footnote{
See Ref.~\cite{Nakayama:2008qz} for a comprehensive survey of
possible choices for $\CN=3$ Chern-Simons matter theories.}
For the ABJM model, there is a natural global $U(1)$ symmetry
under which all fields $Z_\a, \Psi^\b$ carry the same charge and
here we are interested in keeping only particles once we
identify the particle number as this global $U(1)$ charge.

As we are looking at the low energy dynamics in the symmetric phase,
let us begin by the scalar Lagrangian, ignoring the higher-order
interaction terms for a while
\begin{eqnarray}
  \CL_\text{scalar} =
  \frac{1}{c^2} D_t \bar{Z}^\a D_t  Z_\a  -
  D_i \bar{Z}^\a D_i  Z_\a - \frac{m^2c^2 }{\hbar^2} \bar{Z}^\a Z_\a\ .
\end{eqnarray}
Considering the particle modes in the scalar fields
\begin{eqnarray}
  Z_\a = \frac{\hbar}{\sqrt{2m}} \ z_\a e^{-imc^2t/\hbar }\ ,
\end{eqnarray}
the above Lagrangian in the nonrelativistic limit $c\to \infty$
becomes
\begin{eqnarray}
  \CL_\text{scalar}^\text{NR} = \bar{z}^\a \Big(
  i\hbar  D_t  + \frac{\hbar^2 }{2m} D_i^2 \Big)
  z_\a\ ,
\end{eqnarray}
where $A_t = cA_0$ is kept finite. The correction term $\sim \hbar^2 |\p_t z_\a|^2/mc^2 $
is smaller than the above Lagrangian by the factor  $\big(p /mc\big)^2 $.
As will be explained later, there would be in addition non-vanishing
contributions from the quartic interaction terms.
The Chern-Simons term in the nonrelativistic limit becomes
\begin{eqnarray}
  \CL_\text{CS} = \frac{ k  \hbar }{4\pi }\e^{\mu\nu\rho}
  \text{Tr} \big( A_\mu \partial_\nu A_\rho -  \frac{2i}{3}
  A_\mu A_\nu A_\rho  - \tilde{A}_\mu \partial_\nu \tilde{A}_\rho
  +  \frac{2i}{3}  \tilde{A}_\mu \tilde{A}_\nu \tilde{A}_\rho \big),
\end{eqnarray}
where $\mu,\nu,\rho$ runs $t, x^1,x^2$  instead of
$x^0=ct, x^1, x^2$.

For the fermionic part of the Lagrangian without the Yukawa
interaction
\begin{eqnarray}
  \CL_\text{fermion} &=&  -i\bar{\Psi}  \gamma^\mu D_\mu \Psi
  \mp \frac{i m c}{\hbar} \bar{\Psi} \Psi \nn \\
  &=& -i  \Psi^\dagger \gamma^0 \Big(
  \frac{1 }{c}\g^0 D_t +   \g^i  D_i   \Psi
  \pm  \frac{mc}{\hbar}  \Big)  \Psi\ ,
\end{eqnarray}
it needs a little more elaboration to take the nonrelativistic limit.
The upper sign for the mass is for $\Psi^1, \Psi^2$ and the lower sign for $\Psi^3,\Psi^4$. We choose the three-dimensional gamma matrices
to be
\begin{eqnarray}
  \g^0=i\tau^2\ , \qquad
  \g^1=\tau^1\ , \qquad
  \g^2=\tau^3\ , \qquad \g^{012} = 1\ .
\end{eqnarray}
Keeping only the particles again, one can expand the fermion fields
as
\begin{eqnarray}
  \Psi (t,{\bf x}) =  \sqrt{ \hbar c} \Big(
  u_+ \psi_- (t,{\bf x}) + u_- \psi_+ (t,{\bf x}) \Big)
  e^{-imc^2t/\hbar}\ ,
\end{eqnarray}
where $\psi_\pm$ are single-component Grassmann fields
and $u_\pm$ are orthonormal two-component constant spinors such that 
\begin{eqnarray}
  u_{\pm} =\frac{1}{\sqrt{2}} \left( \begin{array}{c} 1 \\ \mp i \end{array}\right)\ ,
  \qquad
  u_+^\dagger u_-=1\ , \qquad
  u^\dagger_+u_+=0\ .
\end{eqnarray}
Since these constant spinors $u_\pm$ carry spin $\pm 1/2$,
\begin{eqnarray}
  -\frac i2 \g^{12} \cdot  u_\pm = \pm \frac12 u_\pm  \nn \ ,
\end{eqnarray}
$\psi_\mp$ annihilates  a particle of spin $\pm 1/2$,
and $(\psi_\mp)^\dagger = \bar\psi_\pm $ creates one.
Defining  $D_\pm = D_1\pm  i D_2$ and $A_\pm = A_1\pm i A_2$,
the fermionic Lagrangian can be rewritten as
\begin{eqnarray}
  \CL_\text{fermion} &=&  \hbar c\bar\psi_+
  \Big( \frac{i}{c}D_t \psi_-  +\frac{mc}{\hbar}(1\mp 1) \psi_- - iD_- \psi_+  \Big)
  \nn \\  && \hspace*{0.3cm}
  + \hbar c \bar\psi_- \Big(\frac{i}{c} D_t \psi_+ + \frac{mc}{\hbar}(1\pm 1)\psi_+
  - iD_+\psi_- \Big)\ .
\end{eqnarray}
Using the equation of motion for $\bar \psi  $ up to the leading order,
one can show that one of the components $\psi_\pm$ is completely determined by the other
\begin{eqnarray}
  \left\{ \begin{array}{ll}
  \psi_+ = \frac{i\hbar}{2mc} D_+ \psi_-  - \frac{i\hbar}{2mc^2} D_t \psi_+
  & \text{for upper sign}\ , \\
  & \\
  \psi_- = \frac{i\hbar}{2mc} D_- \psi_+ -   \frac{i\hbar}{2mc^2} D_t \psi_-
  & \text{for lower  sign}\ . \end{array} \right.
  \label{nonrelf}
\end{eqnarray}
It implies that the spin of dynamical modes in the nonrelativistic limit
is correlated with the sign of the mass.
The correction from the Yukawa interaction to (\ref{nonrelf})
is again of order $1/c^2$ and therefore negligible.
Inserting the above relations, the nonrelativistic fermionic
Lagrangian becomes
\begin{eqnarray}
  \CL_\text{fermion}^{NR}
  = \left\{ \begin{array}{ll}
  \bar\psi_+  ( i\hbar D_t + \frac{\hbar^2 }{2m}D_-D_+ )\psi_-
  + {\cal O}(\frac{1}{c}) & \text{ for upper sign}\ , \\ & \\
  \bar\psi_- ( i\hbar D_t + \frac{\hbar^2}{2m} D_+ D_- )\psi_+
  + {\cal O}(\frac{1}{c}) & \text{ for lower sign}\ .
  \end{array} \right.
\end{eqnarray}
There would be additional contributions from the Yukawa interaction terms
to the above Lagrangian as presented in the next subsection.

\subsection{The nonrelativistic superconformal  ABJM model}

The nonrelativistic limit of the ABJM model for the $U(N)\times U(N)$ gauge group  is
made of many parts. We now have to take into account the higher order
interaction terms. The scalar potential contains quadratic mass terms,
negative quartic terms, and positive sextic terms.
In nonrelativistic limit, the bosonic part of the full Lagrangian
becomes
\begin{eqnarray}
  \CL_\text{scalar}  =  \text{Tr} \Big[  i\hbar  \bar{z}^\a D_t z_\a
  - \frac{\hbar^2}{2m}  D_i \bar{z}^\a D_i z_\a
  - \frac{\pi \hbar^2 }{mk  } (z_a\bar{z}^\a z_\b \bar{z}^\g \Omega^\b_{\ \g}
  -\bar{z}^\a z_\a  \bar{z}^\b z_\g \Omega^\g_{\ b} )\Big]\ ,
\end{eqnarray}
where $\Omega^\a_{\ \b} = {\rm diag}(1,1,-1,-1)$.
The sextic interaction terms vanish in the large $c$ limit.
In terms of the dynamical fermion modes, denoted by
one-component anticommuting variables
\begin{eqnarray}
  \psi^\a = \big(\psi^1_-,\psi^2_-,\psi^3_+, \psi^4_+ \big)\ , \qquad
  \bar{\psi}_\a = \big(\psi^\a\big)^\dagger =
  \big(\bar{\psi}_{+1},\bar{\psi}_{+2},\bar{\psi}_{-3},\bar{\psi}_{-4} \big)\ ,
\end{eqnarray}
the kinetic terms for fermions can be expressed as
\begin{eqnarray}
  \CL_\text{fermion}
  = \text{Tr} \Big[i\hbar \bar{\psi}_\a  D_t \psi^\a - \frac{\hbar^2}{2m} D_i
  \bar{\psi}_\a D_i \psi^\a +\frac{\hbar^2}{2m} \Omega^\a_{\ \b}
  \big(\bar{\psi}_\a F_{12} \psi^\b  -\tilde{F}_{12}\bar{\psi}_\a \psi^\b \big)\Big]\ ,
\end{eqnarray}
where the last two terms are Pauli interactions.
Since the mass deformation breaks the $SU(4)$ symmetry, it is convenient
to decompose $SU(4)$ indices into two $SU(2)$ indices $z_\a = (z_a, z_i)$
where $a=1,2$ and $i=3,4$.
%Note that
%%
%\be \psi^1_+,\psi^2_+,\psi^3_-,\psi^4_- \sim {\cal O}( \frac{p}{mc}) \psi^\a, \ee
%%
%and so are negligible.
The Yukawa coupling in the $c\to \infty$ limit can then be
expressed as
\begin{eqnarray}
  \CL_\text{Yukawa} &=& \frac{\pi  \hbar^2 }{mk } \text{Tr}
  \Big[ \bar{z}^\a z_\a \big(\bar\psi_a\psi^a - \bar\psi_i\psi^i\big)
  + z_\a\bar z^\a \big(\psi^a\bar\psi_a - \psi^i \bar\psi_i\big)  \nn \\
  && -2 \big(z_a\bar{z}^b \psi^a\bar\psi_b
  + \bar{z}^az_b \bar\psi_a \psi^b \big)
  + 2 \big(z_i\bar{z}^j \psi^i\bar\psi_j +\bar{z}^i z_j \bar\psi_i \psi^j \big)
  \nn \\
  && -2 \e_{ab}\e_{ij} \big( \bar{z}^a\psi^b\bar{z}^i\psi^j
  +\bar{z}^a \psi^i \bar{z}^j \psi^b\big) -2 \e^{ab}\e^{ij}
  \big(z_a\bar{\psi}_b z_i \bar{\psi}_j + z_a\bar\psi_i z_j \bar\psi_b\big)
  \Big]\ .
\end{eqnarray}
As a result, the nonrelativistic ABJM Lagrangian can be written as
the sum of the kinetic part and the potential part.
\begin{eqnarray}\label{faction}
  \CL_\text{NR} &=&
  \frac{ k \hbar }{4\pi }\e^{\mu\nu\rho}
  \text{Tr} \Big( A_\mu \partial_\nu A_\rho -  \frac{2i}{3}
  A_\mu A_\nu A_\rho  - \tilde{A}_\mu \partial_\nu \tilde{A}_\rho
  + \frac{2i}{3}  \tilde{A}_\mu \tilde{A}_\nu \tilde{A}_\rho \Big)
  \nn \\ &&
  +  i\hbar  \text{Tr} \Big(\bar{z}^\a D_t z_\a + \bar\psi_\a D_t \psi^\a \Big)
  -\hbar^2 \CH
\end{eqnarray}
with the Hamiltonian density $\hbar^2 \CH$
\begin{eqnarray} \label{action}
  \CH &=& \frac{1}{2m} \text{Tr} \Big(
  D_i \bar z^\a D_i z_\a + D_i \bar\psi_\a D_i \psi^\a \Big)
  -\frac{1}{2m}\Omega^\a_{\ \b} \text{Tr} \Big(\bar{\psi}_\a F_{12}\psi^\b
  - \tilde{F}_{12}\bar\psi_\a \psi^\b \Big)
  \nn \\ &&
  + \frac{\pi}{mk} \text{Tr}
  \Big(z_\a\bar{z}^\a \big(z_\b \bar{z}^\g - \psi^\g \bar\psi_\b \big) \Omega^\b_{\ \g}
  -\bar{z}^\a z_\a  \big( \bar{z}^\b z_\g + \bar\psi_\g \psi^\b) \Omega^\g_{\ \b} \big)
  \nn \\ &&
  + \frac{2\pi}{mk} \text{Tr} \Big[
  \big(z_a\bar{z}^b \psi^a\bar\psi_b  + \bar{z}^az_b \bar\psi_a \psi^b \big)
  - \big(z_i\bar{z}^j \psi^i\bar\psi_j +\bar{z}^i z_j \bar\psi_i \psi^j \big)
  \nn \\ &&
  + \e_{ab}\e_{ij} \big(\bar{z}^a\psi^b\bar{z}^i\psi^j
  + \bar{z}^a \psi^i \bar{z}^j \psi^b \big)
  + \e^{ab}\e^{ij} \big(z_a\bar{\psi}_b z_i \bar{\psi}_j
  + z_a\bar\psi_i z_j \bar\psi_b\big) \Big]\ .
  \label{hamilton}
\end{eqnarray}
The Gauss law constraints for the two gauge groups $U(N)\times U(N)$
are
\begin{eqnarray}
  {\cal G} &\equiv&  F_{12}  +\frac{2\pi}{k}
  \big(z_\a \bar z^\a - \psi^\a \bar\psi_\a \big) ~=~ 0\ ,
  \nn \\
  \tilde{\cal G} &\equiv&  \tilde{F}_{12} + \frac{2\pi}{k }
  \big(\bar{z}^\a z_\a +\bar\psi_\a \psi^\a \big) ~=~ 0\  .
\label{gauss}
\end{eqnarray}
The Lagrangian can be rewritten as
\begin{eqnarray}
  \CL_\text{NR} &=&
  \frac{k\hbar}{2\pi} \tr (A_2 \p_t A_1 - \tilde{A}_2 \p_t \tilde{A}_1)
  + \frac{k\hbar}{2\pi}\tr (A_t {\cal G}- \tilde{A}_t \tilde{\cal G})
  \nn \\ &&
  +i\hbar \tr (\bar z^\a \p_t z_\a + \bar\psi_\a \p_t \psi^\a)  -\hbar^2 {\cal H}\ .
\end{eqnarray}
The quartic potential term in the Hamiltonian is negative.
It is well-known that such a term leads to attraction among particles. However, the total Hamiltonian density can be shown to be positive definite once we impose the Gauss law constraints.
This is consistent with the superalgebra
$\int d^2 x \, \CH \sim \{ Q, Q^\dagger\} $.

Let us now in turn discuss the quantization of the present model. The
equal-time canonical commutation relations can be read off from (\ref{faction})
\begin{eqnarray}
 \big[ A_+(x)^A_{\ B} , A_-(y)^C_{\ D} \big] &=&
 +\frac{4\pi }{k} \d^A_D\d^C_B \d^2(x-y)\  , \nn \\
 \big[ \tilde{A}_+(x)^M_{\ N} , \tilde{A}_-(y)^P_{\ Q}\big] &=&
 -\frac{4\pi }{k} \d^M_Q \d^P_N \d^2(x-y)\ , \nn  \\
 \big[z_\a(x)^A_{\ M}, \hspace*{0.2cm}\bar z^\b(y)^N_{\ B}\big] &=&
 +\d^\b_\a \d^A_B \d^N_M  \d^2(x-y)\  , \nn \\
 \big\{ \psi^\a(x)^A_{\ M}, \bar\psi_\b(y)^N_{\ B} \big\}
 &=&
 +\d^\a_\b \d^A_B \d^N_M \d^2(x -y)\ ,
\end{eqnarray}
where the gauge group indices are included for completeness.
Any physical state $|\Psi\rangle $ is required to satisfy the Gauss law constraints
\begin{eqnarray}
  \CG|\Psi\rangle = \tilde \CG|\Psi \rangle = 0\ .
\end{eqnarray}
To define the physical Hilbert space with the positive Chern-Simons level $k>0$,
let us introduce the Fock  vacuum $| \Omega \rangle$ such that
\begin{eqnarray}
  A_-(x) |\Omega \rangle = \tilde{A}_+(x)| \Omega \rangle
  =z_\a(x)| \Omega \rangle = \psi^\a(x)| \Omega \rangle  = 0\ .
\end{eqnarray}
Obviously, the state $| \Omega \rangle$ does not satisfy the Gauss law
constraints and can not define the physical vacuum.
Instead, the physical vacuum would be defined as
\begin{eqnarray}
  |0\rangle  = \CU(A_+, \tilde A_-) |\Omega \rangle
\end{eqnarray}
with a certain functional $\CU$ such that
\begin{eqnarray}
  F_{+-}|0\rangle = \tilde{F}_{+-}|0\rangle = 0 \ .
\end{eqnarray}
Without matter fields,
the properties of the functional $\CU(A_+, \tilde A_-)$, called the cocycle factor, in the Schr\"odinger picture have been studied in \cite{Dunne:1989cz,Elitzur:1989nr}.
The excited states obtained by matter creation operators should
modify the corresponding  operator $\CU$ so that the configuration still
satisfies the Gauss law. It is not easy to solve for $\CU$
for all excited states.

The Gauss laws dictates that particles created by operators $\bar z^\a, \bar\psi_\a$
should be accompanied by nonabelian magnetic flux.
Each particle with nonabelian charges should be dressed by
nonabelian flux and their creation/annihilation operators create/annihilate
both charge and flux at the same time.
We thus expect  the existence of a dressed operator for each charged field, say
\begin{eqnarray}
  z_\a(x)^A_M, \psi^\a(x)^A_{\ M}  \Longrightarrow
  \CZ_a(x)^A_{\ M},  \varPsi^\a(x)^A_{\ M}\ .
\end{eqnarray}
Note that $\CZ_\a, \varPsi^\a$ and their conjugates commute with the generators of
the local gauge transformation $\CG,\tilde{\CG}$ but transform as $N\times \bar N$
under the global part of the gauge group $U(N)\times U(N)$.  These operators would
annihilate and create physical particles, and are quasi-local
in the sense the magnetic flux are fractional and so detectible in large distance.
For the $U(1)\times U(1)$ ABJM model, one can explicitly
construct such dressed operators: introducing a dual photon $\s$
of the field strength $F+\tilde F$, the dressed operators can be expressed as
\begin{eqnarray}
  \CZ_\a = e^{i\s/k} z_a, \qquad  \varPsi^\a = e^{i\s/k} \psi^\a\ ,
\end{eqnarray}
when we normalize the dual photon $\s$ to be $2\pi$ periodic.

\section{Symmetries, Conserved Charges and their Algebra}

We have now a specific nonrelativistic Lagrangian for particles in the symmetric phase
of the mass-deformed ABJM model. Not only has it inherited internal symmetries from
the relativistic theory, it also has the nonrelativistic limit of spacetime symmetries and supersymmetries. We present in this section the symmetry group respected by
the present nonrelativistic ABJM model. It turns out that the model
has three-dimensional super Schr\"odinger group with fourteen supercharges.

For a technical comment, we need to be careful about the operator ordering
for the density of conserved charges. Besides the Hamiltonian,
all densities are quadratic and are normal-ordered.
The algebra fixes almost everything. Hereafter we put the Plank constant
$\hbar=1$ for simplicity.

\subsection{Internal symmetry}

The original ABJM theory has the $SU(4)$ R-symmetry and $U(1)$ global symmetry.
We are keeping only particles with respect to this $U(1)$ global symmetry
under which the particles, annihilated by canonical fields $z_\a, \psi^\a$,
have the unit charge. In the nonrelativistic theory, this global
$U(1)$ charge can be therefore identified as the particle number operator,
which takes the form
\begin{eqnarray}
  \CN = \int d^2x \,  n(x)\ ,
\end{eqnarray}
where the number density is given by
\begin{eqnarray}
  n(x) = \text{Tr} \Big(
  \bar z^a z_a + \bar z^i z_i + \bar \psi_a \psi^a
  + \bar\psi_i \psi^i \Big)\ .
\end{eqnarray}
It is sometimes useful to define the total mass operator $\CM = m \CN$.

The mass deformation reduces the $SU(4)$ R-symmetry down to
$SU(2)_L\times SU(2)_R\times U(1)_R $. As seen in (\ref{faction},\ref{action}),
the nonrelativistic limit does not violate any of these symmetries.
The fundamental fields transform under the R-symmetry inherited
from the mother theory as
\begin{eqnarray}
  z_a: \big({\bf 2},{\bf 1}\big)_{1/2}\ , \ \
  z_i: \big({\bf 1},{\bf 2}\big)_{-1/2}\ , \ \
  \psi^a: \big({\bf \bar 2},{\bf 1}\big)_{-1/2}\ , \ \
  \psi^i: \big({\bf 1},{\bf \bar 2}\big)_{1/2}\ .
\end{eqnarray}
Their N\"other charges are given by
\begin{eqnarray}
  R^a_{\ b} &=& - \int d^2x\ \text{Tr}
  \Big[  ( \bar{z}^a z_b  - \bar\psi_b  \psi^a   ) -
  \frac12 \d^a_{\ b} \big(\bar z^c z_c - \bar\psi_c \psi^c \big) \Big]\ ,
  \\
  R^j_{\ i} &=& - \int d^2 x \
  \text{Tr} \Big[ \big(\bar{z}^j   z_i  - \bar\psi_i  \psi^j \big) - \frac12 \d^j_{\ i}
  \big(\bar z^k z_k -\bar \psi_k \psi^k \big) \Big]
\end{eqnarray}
for $SU(2)_L$ and $SU(2)_R$ and
\begin{eqnarray}
  R = \frac12 \int d^2x\  \text{Tr} \Big(
  \bar{z}^a z_a  + \bar\psi_{-i}\psi^i_+ -\bar z^i z_i -\bar\psi_{+a}\psi^a_-
  \Big)
\end{eqnarray}
for $U(1)_R$.  There is actually an additional $U(1)$
symmetry that arises in the nonrelativistic limit.
As presented in the previous section, massive fermions
in the nonrelativistic limit carry specific spin
values $\pm 1/2$ depending on the sign of mass terms.
In the nonrelativistic theory, the sum of fermion
spin is also conserved by itself.
The total spin of massive fermions can be expressed as
\begin{eqnarray}
  \Sigma  =  \frac12 \int d^2x \, \text{Tr}\Big(
  \bar\psi_{+a}\psi^a_- - \bar\psi_{-i}\psi^i_+\Big)\ .
\end{eqnarray}
The charges of creation operators or particles for these internal symmetries
are summarized in the Table \ref{gcharges0}.
\begin{table}
\caption{Internal charges for the creation operators}
\begin{center}
\begin{tabular}{|c|cccc|}
  \hline
  & $\bar z^a$ & $  \bar\psi_a$ & $\bar z^i $ &  $ \bar{\psi}_i  $ \\
  \hline $\CN$  &   $ 1$ & $  1$ & $ 1$ & $  1$ \\
  $R$ & $1/2$ & $- 1/2$ & $-1/2 $ & $1/2$ \\
  $\Sigma$ & 0 & $1/2 $ & 0 & $-1/2 $ \\
  $SU(2)_1$& ${\bf 2} $ & $\bar{\bf 2}$ & 1 & 1 \\
  $SU(2)_2 $ & 1 & 1 & ${\bf 2}$ & $\bar{\bf 2}$ \\
\hline
\end{tabular}
\end{center}
\label{gcharges0}
\end{table}

\subsection{Space-time symmetry}

The Poincar\'e symmetry is reduced to the Galilean symmetry
in the nonrelativistic limit. They are generated by
the Hamiltonian $H$, momenta $P_i$, rotation $J$
and Galilean boosts $G_i$. The time and space translational
symmetry lead to the conserved Hamiltonian $H$ and
linear momentum $P_i$. With the Hamiltonian density $\CH$ in Eq.(\ref{hamilton}),
the conserved Hamiltonian and linear momentum are given as
\begin{eqnarray}
  H = \int d^2 x\ \CH , \qquad  P_i = \int d^2x \
  \CP_i\ ,
\end{eqnarray}
where the momentum density is
\begin{eqnarray}
  {\cal P}_i  = -\frac{i}{2} \text{Tr}\Big[
  \bar{z}^a D_i z_\a - \big(D_i \bar{z}^\a \big) z_\a
  +\bar{\psi}^\a D_i \psi-D_i\bar{\psi}^\a \psi_\a \Big]\ .
\end{eqnarray}
The rotational symmetry with transformation $\d x_i = \a \epsilon^{ij}x_j$
leads to the conserved angular momentum.
The angular momentum in the nonrelativistic limit takes the form
\begin{eqnarray}
  J = \int d^2 x \
  \big( x_1 \CP_2- x_2 \CP_1 \big)  + \Sigma\ ,
\end{eqnarray}
sum of the orbital and spin angular momentum.
The Lorentz boosts in the nonrelativistic limit
reduce to the Galilean boosts
$ \d t = 0 , \  \d x_i  = \a_i  t $,
whose conserved charges are
\begin{eqnarray}
  G_i = -t P_i  +m \int d^2x \
  x_i\ n(x)\  .
\end{eqnarray}
They are related to the position of the center of mass of the whole system.

Even though the mass-deformation breaks the conformal symmetry
of the relativistic ABJM model, the nonrelativistic limit introduces
a new kind of the nonrelativistic conformal symmetry,
called the Schr\"odinger symmetry. Because the nonrelativistic
ABJM model has quartic interactions only, the Lagrangian
is invariant under a dilatation symmetry with transformation
$\d t= \a^2 t, \ \d x_i = \a x_i $ and canonical
scale transformations for the matter fields.
The conserved charge is
\begin{eqnarray}
  D = 2H t -  \int d^2x\ x_i \CP_i\  .
\end{eqnarray}
In addition to this scale symmetry, the nonrelativistic
ABJM model also preserves a single special conformal
symmetry with transformation $\d t = \a t^2, \ \d x_i  = -\a t x_i $
whose charge can be expressed as
\begin{eqnarray}
  K = -t^2 H +t D + \frac{m}{2} \int d^2x \
  x_i^2\,  n(x)\ .
\end{eqnarray}

Let us then present the algebra these generators should satisfy.
The generators $H, P_\pm= P_1\pm i P_2, J, G_\pm=G_1\pm i G_2 $
satisfy the Galilean algebra with the particle number $\CN$ as a central term
\begin{eqnarray}
  && [H, P_\pm ]=0\ , \ \ [ P_+, P_-]=0\ , \ \ [J,H]=0\ ,\ \  [J,H]=0\ , \nn \\
  && [ J, P_\pm  ]= \pm  P_\pm\ , \ \ [G_+, G_-] =0\ , \ \
  [J, G_\pm ]=   \pm  G_\pm\  , \ \nn \\
  &&  i[H, G_\pm ]= P_\pm\ , \ \
  i[P_+,G_-] = i[P_-,G_+] = 2m \CN\  .
\end{eqnarray}
It is noteworthy that the $H, D, K$ generates the conformal subalgebra
$SO(2,1)$,
\begin{eqnarray}
  i[D, H]= 2H\ , \ \  i[D, K] = -2K\ , \ \
  i[K, H]= D\ .
\end{eqnarray}
With the additional commutation relations
\begin{eqnarray}
  && [D,J]=0\ , \ \  i[D, P_\pm ]= P_\pm\ , \ \ i[D, G_\pm] = -G_\pm \ ,  \\
  && [K, J]=0\ , \ \  i[K, P_\pm] = -G_\pm\ , \ \ [K ,G_\pm]= 0 \ ,
\end{eqnarray}
these charges for the space-time symmetry generate
the three-dimensional Schr\"odinger (conformal-Galilean)
algebra.
%The scaling dimension $\D_\CO$ of an operator is defined by
%%
%\be i[D,\CO]= \D_\CO \CO . \ee
%%
%One can read the scaling dimension of $z_\a, \psi^\a$ is $1$ and $D,H, P_\pm, K, G_\pm$ are
%$0,2,1,-2,-1$, respectively.

\subsection{Supersymmetry}

An interesting generalization of this Schr\"{o}dinger algebra is to
introduce fermionic conserved charges.
The algebra then can be enhanced to a super
Schr\"{o}dinger algebra. One important feature of the nonrelativistic
supersymmetry is that there are two different types of supercharges:
one is called `dynamical supercharges' $Q_\text{D}$ and
another is called `kinematical supercharges' $q_\text{K}$.
They satisfy roughly the anti-commutation relations
\begin{eqnarray}
  \big\{Q_\text{D}, Q_\text{D}^\dagger \big\} \sim H\ ,
  \qquad
  \big\{ q_\text{K}, q_\text{K}^\dagger \big\} \sim
  \CN\ .
\end{eqnarray}
Compare to relativistic superconformal symmetry, there is relatively
much room to extend the Schr\"{o}dinger algebra
by adding kinematical supercharges
(see, however, discussion in section \ref{comp-s}). We first
therefore have to manifest how the Schr\"{o}dinger algebra
is extended in the NR limit of mass-deformed ABJM model.

We start from the consistent truncation of the
relativistic supersymmetry transformations (\ref{susytrans})
of the fields in the nonrelativistic limit. Expanding
the SUSY parameters $\xi_{\a\b}$ as
\begin{eqnarray}
 \xi_{\a\b}=  \xi_{-\a\b  } u_+ + \xi_{+\a\b  } u_- \ ,
 \qquad \xi_\pm^{\a\b } = (\xi_{\mp\a\b})^\dagger
 = \frac12 \epsilon^{\a\b\g\d}\xi_{\pm \g\d}\ ,
\end{eqnarray}
and using the relations (\ref{nonrelf}) for the non-dynamical
modes $\psi^a_+, \psi^i_-$, the SUSY variation rules
for scalar fields can be expanded as
\begin{eqnarray}
  \d z_a  &=& \sqrt{\frac{2mc}{\hbar}} \big(
  \xi_{+ab} \psi^b_-  -\xi_{-ai} \psi^i_+ \big)
  + \sqrt{\frac{\hbar}{2mc}}\big(- \xi_{-ab}  i D_+\psi_-^b
  + \xi_{+ai} i  D_-\psi_+^i  \big)\ ,  \\
  \d z_i  &=&  \sqrt{\frac{2mc}{\hbar}}\big( -\xi_{-ij} \psi^j_+
  + \xi_{+ia}  \psi^a_- \big) +  \sqrt{\frac{\hbar}{2mc}}
  \big( \xi_{+ij}  i D_- \psi_+^b  - \xi_{-ia}i  D_+\psi_-^a  \big)\ ,
\end{eqnarray}
up to the second leading orders.
%Note that $\xi_{ab}= -\xi_{12}\epsilon_{ab}$
%and $\xi_{ij}=-\xi_{34}\epsilon_{ij}$ with $\e_{12}= \e_{34} =-1$.
As mentioned before one can see there are two kinds of transformation
rules, of which one is the leading-order terms
\begin{eqnarray}
  \d_K z_a = \xi_{+ab} \psi^b_-  - \xi_{-ai}  \psi^i_+ \ , \qquad
  \d_K z_i =  - \xi_{- ij}  \psi^j _+ + \xi_{+ia}\psi^a_-\ ,
\end{eqnarray}
and the other is the next-to-leading order terms
\begin{eqnarray}
  \d_D z_a = \frac{-i}{2m}  \xi_{-ab}   D_+ \psi^b_-\ , \qquad
  \d_D z_i = \frac{i}{2m} \xi_{+ij}   D_- \psi^j_+ \ .
\end{eqnarray}
The former will be identified with the kinematical supersymmetry and the latter with the `dynamical supersymmetry. Here we rescaled the
supersymmetry parameters as
\begin{eqnarray}
  \sqrt{\frac{2mc}{\hbar}}\big( \xi_{+12} , \xi_{-34},
  \xi_{\pm ai} \big) &\rightarrow& \big( \xi_{+12} ,  \xi_{-34} ,\xi_{\pm ai}\big)
  \ , \nn \\
  \sqrt{\frac{2m\hbar}{ c} } \big(\xi_{-12},\xi_{+34}\big)
  &\rightarrow&  \big(\xi_{-12}, \xi_{+34}\big)\ ,
\label{scale}
\end{eqnarray}
to keep the transformation rules finite. Note that the
total angular momentum is manifestly preserved on the right-hand side
of the above transformation rules.
%The nonrelativistic supersymmetry of the scalar field is made of
%two kinds of the transformation. The first one is the
%leading order  supersymmetries with transformation
% are
% %
% \be && \d_K z_a = \xi_{+ab} \psi^b_-  - \xi_{-ai}  \psi^i_+ , \ \
%  \d_K z_i =  - \xi_{- ij}  \psi^j _+ + \xi_{+ia}\psi^a_- .  \ee
% %
% The next leading order transformation is that for the dynamical supersymmetry,
% %
% \be && \d_D z_a = \frac{-i}{2m}  \xi_{-ab}   D_+ \psi^b_- , \ \
%  \d_D z_i = \frac{i}{2m} \xi_{+ij}   D_- \psi^j_+  .\ee
% %
% Note that the spin of $D_\pm$ is $\pm 1$ and so the total spin is preserved.

One can also work out the nonrelativistic limit of the fermionic supersymmetry
transformations. Applying the same idea, one
can read off the leading order kinematical
supersymmetry transformation rules
\begin{eqnarray}
\d_K\psi^a_- = \xi^{ab}_- z_b + \xi^{ai}_-  z_i\  ,  \qquad
  \d_K \psi^i_+  = -\xi^{ij}_+  z_j -\xi^{ia}_+  z_a\ .
\end{eqnarray}
The subleading dynamical supersymmetry transformation rules becomes
\begin{eqnarray}
  && \d_D \psi^a_- =  \frac{-i}{2m}\xi_+^{ab}  D_- z_b\ , \qquad
  \d_D \psi^i_+ = \frac{i}{2m} \xi^{ij}_- D_+ z_j\ .
\end{eqnarray}
It is interesting to note that there are no other contributions
to the dynamical supersymmetry transformation rules
expect the canonical terms even in the interacting theory.

Let us now in turn consider the transformation of the gauge field.
The kinematical and dynamical supersymmetry transformations for $A_t$ are
\begin{eqnarray}
 \d_K A_t &=& -\frac{\pi}{mk }\Big( z_a (\bar\psi_{+b}\xi^{ab}_-
 + \bar\psi_{-i}\xi^{ai}_+)
 + \psi^a_-(\xi_{+ab}\bar z^b + \xi_{+ai}\bar z^i ) \nn \\
 && \hspace*{1.1cm} + z_i(\bar\psi_{-j}\xi^{ij}_+  + \bar\psi_{+a}\xi^{ia}_ - )
 + \psi^i_+ ( \bar z^a \xi_{-ia} + \bar z^j \xi_{-ij} )\Big)\ ,  \\
 \d_D A_t &=&  \frac{\pi i}{2m^2 k } \Big( z_a D_-\bar\psi_{+b} \xi^{ab}_+
  - D_+ \psi^a_- \xi_{-ab}\bar z^b + z_i D_+ \bar \psi_{-j} \xi^{ij}_-
 -D_- \psi^i_+ \xi_{+ ij} \bar z^j \Big)\ .
\end{eqnarray}
For the spatial part of the gauge field $A_\pm = A_1\pm  i A_2$,
the kinematical and dynamical supersymmetry transformations are
\be && \d_K A_\pm =0 ,    \\
  && \d_D A_-  = \frac{2\pi}{mk }(\psi_-^a\bar z^b \xi_{-ab} + z_i \bar\psi_{-j}\xi^{ij}_-  ), \nn  \\
&& \d_D A_+ = \frac{2\pi}{mk } (\psi^i_+\bar z^j \xi_{+ij} + z_a \bar\psi_{+b} \xi^{ab}_+ ) . \ee
Looking at the kinematical and dynamical supersymmetry variation rules,
one can conclude that the twelve SUSY parameters split into
kinematical and dynamical ones as
\begin{eqnarray}
  \big\{ \eta_{\pm 12}, \eta_{\pm 34} , \eta_{\pm ai} \big\}
  \Longrightarrow
  \underbrace{(\eta_{-12}, \eta_{+34}, \eta_{\pm ai})}_\text{kinematical}
  + \underbrace{(\eta_{+12}, \eta_{-34})}_\text{dynamical}\ ,
\end{eqnarray}
{\it i.e.}, the nonrelativistic ABJM model has
ten kinematical supercharges and two dynamical supercharges.

%there would be conserved
%supercharges. The infinitesimal supersymmetric transformation is generated
%anti-hermitian generator $\eta_{+\a\b}Q^{\a\b}_- /2+ \eta_{-\a\b}Q^{\a\b}_+/2$ with
%$\d z_\g =  [\eta_{\pm \a\b} Q^{\a\b}_\mp , z_\g]/2$.
%We note that $\epsilon^{12}=-\epsilon_{12}=1$ and $\epsilon^{34}=-\epsilon_{34}=1$.
%Then $\xi_{ab} = -\xi_{12}\epsilon_{ab}$ and $\xi^{ab}= \xi^{12} \epsilon^{ab}$ and so on.
Given the above transformation rules, one can construct the
conserved kinematical supercharges as
\begin{eqnarray}
 q_- &\equiv&  Q_{K -}^{\ 12} =Q_{K - 34} =  \int d^2  {\bf x}\
 \text{Tr} \Big(+\epsilon_{ab} \bar{z}^a\psi^b_-  -
 \epsilon^{ij} \bar \psi_{-i} z_j  \Big)\ , \nn  \\
 \bar q_+ &\equiv&  Q_{K +}^{\ 34}= Q_{K +12} =
 \int d^2 {\bf x} \  \text{Tr} \Big(-\epsilon_{ij} \bar z^i\psi^j_+   + \epsilon^{ab}
 \bar\psi_{+a} z_b  \Big)\  ,\\
 q^{\ ai}_- &\equiv& \e^{ab}\e^{ij}~ Q_{K -jb}
 =  \int d^2{\bf x} \ \text{Tr} \Big(
 \bar{z}^i \psi^a_- - \epsilon^{ab}\epsilon^{ij} \bar{\psi}_{-j} z_b \Big), \nn \\
 q^{\ ai}_{ +} &\equiv&  \e^{ab}\e^{ij} ~ Q_{K+ ib}
 = \int d^2 {\bf x} \ \text{Tr} \Big( \bar z^a \psi^i_+ - \e^{ab} \e^{ij}   
 \bar\psi_{+b} z_j  \Big)\  , 
\end{eqnarray}
with the relations $q_-^\dagger = \bar q_+$, $q^{\ ai}_{ \pm} = -q^{\ ia}_{ \pm}$,
$(q^{\ 13}_{ -})^\dagger  = q_{+ 13}= -q^{\ 24}_{ +}$, and
the conserved dynamical supercharges as
\begin{eqnarray}
  Q_+ &\equiv& Q_{D+}^{12} = Q_{D+34}
  = \frac{i}{2m} \int d^2 {\bf x}\ \text{Tr} \Big(
  - \epsilon_{ab} \bar z^a  D_+ \psi^b_- + \epsilon^{ij} \bar\psi_{-i} D_+ z_j
  \Big)\ , \nn  \\
  \bar Q_-  &\equiv&  Q_{D-}^{34}=Q_{D-12} = \frac{i}{2m}
  \int d^2 {\bf x}\ \text{Tr} \Big( \epsilon_{ij}   \bar  z^i D_- \psi^j_+
  - \epsilon^{ab}  \bar \psi_{+a} D_- z_b  \Big)\ .
\end{eqnarray}
In addition to these manifest supercharges, one has in the NR ABJM model
another set of conserved fermionic charges, say conformal supercharges $S$.
They arise in the commutator of the special conformal generator
$K$ and the dynamical supercharges $Q$,
\begin{eqnarray}\label{SSch2}
  i \big[ K , Q_+\big] ~\equiv~ + S_+\ ,
  \qquad
  i \big[ K , \bar Q_- \big] ~\equiv~ + \bar S_-\ .
\end{eqnarray}
Their N\"other charges are thus given by
\begin{eqnarray}
   S_+ &=&
   tQ_+ - \frac12 \int d^2 {\bf x} \  \big(x_1+ix_2\big)
   \big(\e_{ab}\bar{z}^a\psi^b -\e^{ij}\bar\psi_i z_j \big)\ , \nn \\
   S_- &=& t Q_- + \frac12
   \int d^2{\bf x}  \ \big(x_1-i x_2\big)
   \big( \e_{ij} \bar z^i \psi^j  - \e^{ab} \bar\psi_a z_b \big)\  .
\end{eqnarray}

We will present the full structure of the symmetry algebra
in the next subsection.
Here, let us pause to discuss aspects of the
kinematical supercharges $q, q^{ai}$
that are main novelties of our $\CN=6$ super-Schr\"odinger algebra.
They satisfy the following anti-commutation relations.
\begin{eqnarray}
  \big\{ q_-, \bar q_+ \big \} &=& \CN\ ,  \  \\
  \big\{ q_-^{\ ai} , q_{+ bj}, \}  &=&
  \frac12 \delta^a_b \d^i_j \CN - \delta^a_b R^i_{\ j} + \d^i_j R^a_{\ b}  
  \  . \label{susyk}
\end{eqnarray}
The former algebra is simply the familiar fermion oscillator algebra. The latter is essentially the same as
the three-dimensional $\CN=4$ Poincar\'e supersymmetry with a non-central
extension
\begin{eqnarray}\label{ncsusy}
  \big\{ \CQ^{ai} , \CQ^{bj} \big\}
  &=& \e^{ab} \e^{ij} C\g^\mu P_\mu   +
   \e^{ij} C \CR^{ab}  -  \e^{ab} C \CR^{ij} \ ,
\end{eqnarray}
which has recently been discussed in \cite{Lin:2005nh,Agarwal:2008pu}.
In particular, it was shown that the particle spectrum for theories
based on the superalgebra (\ref{ncsusy}) does not allow
any massless particles.  One can therefore choose the rest-frame
in which the above algebra can be reduced to (\ref{susyk})
after identification of the mass $P_0$ to the number charge $\CN$.
The above algebra (\ref{susyk}) is also often referred as the $SU(2|2)$
Lie super-algebra with a noncompact $U(1)$ central extension.

The dynamical supercharges $Q_+, \bar Q_-$ and $S_+, \bar S_-$
with their kinematical pairs $q_-, \bar q_+$ now satisfy the $\CN=2$
three-dimensional super Schr\"odinger algebra whose
commutation relations of interest are
\begin{eqnarray}
  \big\{Q_+, \bar Q_- \big\} = \frac{1}{2m} H\ , &&
  \big\{ Q_+, \bar S_-\big\} = \frac{1}{4m} \Big( D-i \big(J-\frac32 \tilde R \big) \Big)
  \ , \nn \\
  \big\{S_+, \bar S_- \big\}  =\frac{1}{2m} K\ , &&
  \big\{ Q_-, \bar S_+ \big\} =\frac{1}{4m} \Big(   D +i  (J -\frac{3}{2} \tilde R  ) \Big)\ ,
  \label{qsc}
\end{eqnarray}
and
\begin{eqnarray}
  \big\{q_+ , Q_+ \big \} = + \frac{1}{2m} P_+\ , &&
  \big\{\bar q_-, \bar Q_- \big\} = + \frac{1}{2m}P_- \ ,  \nn \\
  \big\{ q_+, S_+ \big\} = - \frac{1}{2m} G_+\ , &&
  \big\{ \bar q_-,  \bar S_-\big\} = - \frac{1}{2m} G_-\ .
\label{k4susy}
\end{eqnarray}
The anti-commutation relations in (\ref{qsc}) involve
an interesting modification of $U(1)_R$ charge to the so-called {\it twisted} $U(1)_{\tilde{R}}$ charge
\begin{eqnarray}
  \tilde{R}  = \frac{1}{3} \int d^2{\bf x}\ (2\bar z^a z_a -2\bar z^i z_i -\bar \psi_a \psi^a +\bar\psi_i\psi^i)\ .
\end{eqnarray}
It implies that the original $U(1)_R$ symmetry is mixed with
the fermion spin $\S$ as
\begin{eqnarray}
  \tilde{R}= \frac23 \big( 2R+ \Sigma \big)\ .
\end{eqnarray}
One can check this modification from the Jacobi identity
of $q^{ai}_\pm, Q_+, S_-$, which leads to the invariance
of $q^{\ qi}_\pm$ under the charge $J-3/2 \tilde{R}$.

\subsection{Super-Schr\"odinger algebra: summary}

We found all possible symmetric generators of our theory.
As discussed briefly in the previous subsection,
they satisfy many layers of the algebraic structures.
For later convenience, we summarize the commutation relations
of our super Schr\"odinger algebra with fourteen supercharges.

\paragraph{Schr\"odinger algebra: ${\bf sch_2^{(0)}}$}
\footnote{
It seems standard practice to denote the bosonic
Schr\"odinger algebra in $d$-dimensions by $\bf{sch}_d$.
We introduce an additional superscript to distinguish
several subalgebras of the full super-Schr\"odinger algebra.}
The (2+1)-dimensional
Schr\"{o}dinger algebra is generated
by the Hamiltonian $H$,
momenta $P_i$, Galilean boosts $ G_i$,
rotation $ J$ and special conformal generator $K$,
which satisfy the following commutation relations
\begin{eqnarray}\label{Sch1}
  i \big[ D, H\big] ~=~ + 2 H, \qquad
  i \big[ D, K\big] &=& -2 K, \qquad
  i \big[ K, H\big] ~=~ + D, \nn \\
  i \big[ D, P_i \big] ~=~ + P_i, \qquad
  i \big[ D, G_i \big] &=& - G_i, \qquad
  i \big[ P_i , G_j \big] ~=~ + \d_{ij}  m\CN,  \\
  i \big[ H, G_i \big] ~=~ + P_i, \qquad
  i \big[ K, P_i \big] &=& - G_i, \qquad
  \big[ J , P_{\pm}/G_{\pm} \big] ~=~ \pm P_{\pm}/G_{\pm}\
  \nn .
\end{eqnarray}
Hereafter we present only nonvanishing commutation relations.
The charges $H,D, K$ form a conformal subalgebra $SO(2,1)$.

\paragraph{super Schr\"odinger algebra: ${\bf sch_2^{(1)}}$ }
Six out of the fourteen supercharges of the nonrelativistic ABJM model
are tightly related to the dynamics of the theory: two kinematical
supercharges $q$, two dynamical supercharges
$Q$ and two conformal supercharges $S$. Adding these fermion charges
leads us to a subalgebra we call ${\bf sch_2^{(1)}}$.
These supercharges satisfy the commutation relations
\begin{eqnarray}
  \big\{ q_-, \bar q_+ \big \} = \CN\ ,
\end{eqnarray}
and
\begin{eqnarray}
  \big\{Q_+, \bar Q_- \big\} = \frac{1}{2m} H\ , &&
  \big\{ Q_+, \bar S_-\big\} = \frac{1}{4m} \Big( D-i \big(J-\frac32 \tilde R \big) \Big)
  \ , \nn \\
  \big\{S_+, \bar S_- \big\}  =\frac{1}{2m} K\ , &&
  \big\{ Q_-, \bar S_+ \big\} =\frac{1}{4m} \Big(   D +i  (J -\frac{3}{2} \tilde R  ) \Big)\ ,
\end{eqnarray}
together with
\begin{eqnarray}
  \big\{q_+ , Q_+ \big \} = + \frac{1}{2m} P_+\ , &&
  \big\{\bar q_-, \bar Q_- \big\} = + \frac{1}{2m}P_- \ ,  \nn \\
  \big\{ q_+, S_+ \big\} = - \frac{1}{2m} G_+\ , &&
  \big\{ \bar q_-,  \bar S_-\big\} = - \frac{1}{2m} G_-\ .
\end{eqnarray}
With the space-time symmetry generators, they satisfy
\begin{eqnarray}
  i[D, Q_\pm ] = + Q_\pm\ , \qquad
  i[K, Q_\pm] = + S_\pm\  , &&
  i[G_\pm, Q_\mp] = -q_\pm\ , \nn \\
  i[D, S_\pm] = - S_\pm\ , \qquad
  i[H, S_\pm ] = - Q_\pm\ , &&
  i[ P_\pm, S_\mp]  = -q_\pm\ \ .
\end{eqnarray}
Other quantum numbers of these six supercharges such as
angular momentum $J$, scale dimension $D$ and
$U(1)$ charges $R$, $\S$ and $\tilde R=2(2R+\S)/3$
are summarized in Table \ref{gcharges1}.
The dynamical and superconformal charges $Q_\pm, S_\pm$
together with $(J-\frac{3}{2}\tilde{R})$ extend the $SO(2,1)$ algebra of $H,D, K$ to the superconformal algebra $OSp(2|1)$.

\begin{table}[t]
\caption{Charge for Supersymmetric generators }
\begin{center}
\begin{tabular}{|c|cccc|c|c|}
 \hline
 & $\D $ & $ J $ & $2R $ &  $  2\Sigma $& $ \tilde R$ & $J-\frac32 \tilde{R} $  \\
 \hline
 $Q_+$  &   $ 1$ & $ +1/2 $  & $ +2$   & $   -1$  & $+1$  & $-1$ \\
  $Q_-$  & $1 $    & $ - 1/2 $  & $ -2 $  & $+1$    & $-1$  & $+1$ \\
  $S_+$  & $-1$    & $+1/2  $  &  $ +2$  & $ -1$ & $+1$ &  $-1$ \\
  $S_- $  & $-1  $  & $-1/2  $   & $-2$     & $+1$   & $-1$ & $+1$ \\
 \hline
  $q_+ $ & 0          & $ +1/2  $ & $ -2$   & $+1$   & $-1$ &  $+2$ \\
  $q_- $   &  0        & $ -1 /2   $  &   $+2$  & $-1$    & $+1$   &  $-2$ \\
  $q^{ai}_+ $ &  0  & $+1  /2   $ &  $0 $   & $+1$   & $+1/3$ & $0 $ \\
  $q^{ai}_-$ & 0     &  $-1 /2 $       & $ 0$   & $-1$   &  $-1/3$&  $0$ \\
\hline
\end{tabular}
\end{center}
\label{gcharges1}
\end{table}

\paragraph{super Schr\"odinger algebra: ${\bf sch_2^{(2)}}$ } We add six
generators $R^a_{\ b},
R_i^{\ j}$ of $SU(2)\times SU(2)$ R-symmetry and eight
kinematic supercharges $q^{ai}_\pm$ to $\bf{sch_2^{(1)}}$ above
to arrive at the full algebra ${\bf sch_2^{(2)}}$.
Under the $SU(2)\times SU(2)$ generators, $q_\pm^{ai}$ transform as $(2,2)$. The important commutation relations are
\be
  && [\tilde{R}, q^{ai}_\pm] = 0 , \ [J, q^{ai}_\pm]= \pm \frac12 q^{ai}_\pm, \\
  && \big\{ q_-^{\ ai}, q_{+bj}  \big\}  =
  \frac12 \delta^a_b \d^i_j \CN - \delta^a_b R^i_{\ j}+ \d^i_j R^a_{\ b}  \ .\ee
As mentioned earlier, $q^{ai}_\pm $ satisfy the $SU(2|2)$
Lie super-algebra with a central extension.

\subsection{Comparison with other nonrelativistic super-algebras \label{comp-s}}

We would like to make a few remarks
on the $\CN=6$ super-Schr\"odinger algebra and compare
it with other algebras in the literature.

First, we note that the subalgebra ${\bf sch_2^{(1)}}$
%(which contains the interesting $OSp(2|1)$ sub-algebra of
%$(H,D,K,Q,S,J-\frac{3}{2}R)$)
is common to all super-Schr\"odinger algebras realized by (maximally supersymmetry preserving) non-relativistic limit
of Chern-Simons theories. In other words, the $\CN=2$ \cite{Leblanc:1992wu}, $\CN=3$ \cite{Nakayama:2008qz} and our $\CN=6$ algebras differ only by the kinematical supercharges.

Second, the pattern of splitting of the supercharges to kinematical and dynamical ones contrasts with that of the nonrelativistic superalgebra  obtained by the DLCQ procedure. Let us first
briefly review the DLCQ procedure. Compactifying the light-cone
coordinate $x^-=x^0 -x^3$,
one can formally truncate the relativistic
four-dimensional $\CN=N$ superconformal field theory in a sector of nonzero $P_-$. The symmetry group of the resulting
three-dimensional theory is generated by the
generators of the four-dimensional theory that commute with $P_-$.

The bosonic generators precisely give the Schr\"odinger algebra.
The $4N$ Poincar\'e supercharges  $Q_\a^i, \bar Q_{\da j}$ ($i,j=1,..,N$)
split into $2N$ kinematical supercharges $Q_K$ and $2N$
dynamical supercharges $Q_D$
\begin{eqnarray}
  Q_\a^i \Longrightarrow \left\{
  \begin{array}{ll}
  Q_+^i & \text{: kinematical supercharge } Q_K^i\\
  Q_-^i & \text{: dynamical supercharge } Q_D^i
  \end{array}
  \right. \ ,
\end{eqnarray}
where $\pm$ denote the weight of $SO(2)$ that rotates $12$-plane.
For the conformal supercharges $S_{\a i}, \bar S_\da^{\ i}$, one
can show that only half of them survive the DLCQ procedure
\begin{eqnarray}
  S_\a^i \Longrightarrow \left\{
  \begin{array}{ll}
  S_+^i & \text{: conformal supercharges } S_D^i\\
  S_-^i & \text{: not allowed !}  \end{array}
  \right. \ .
\end{eqnarray}
It implies that our $\CN=6$ super Schr\"odinger algebra
cannot be embedded simply into any four-dimensional
relativistic superconformal algebra. Only the $\CN=2$
sector $(q,Q,S)$ matches with a four-dimensional $\CN=1$
superconformal algebra via DLCQ \cite{Sakaguchi:2008ku}.
We believe this will have some implication on the gravity dual of our theory.

Third, one can try to take the non-relativistic limit
of the mass deformed BLG model \cite{Hosomichi:2008qk,Gomis:2008cv} to obtain yet
another example of super-Schr\"odinger algebra.
The mass deformed BLG model preserves sixteen supercharges
and $SO(4)\times SO(4)$ $R$-symmetry. So, at first sight,
a bigger super-algebra seems likely to appear.
However, it turns out that the resulting theory
only has the same $\CN=6$ super-algebra as the ABJM model.
Recall that, to take the non-relativistic limit,
it is preferred that the elementary fields are complex.
A choice of complex structure of the mass deformed BLG
model breaks the $SO(4)\times SO(4)$ $R$-symmetry
down to $SU(2)\times SU(2)\times U(1) \times U(1)$.
So, it differs from the ABJM model only by an
extra $U(1)$ and extra super-charges charged under that $U(1)$.
A careful analysis shows that these extra super-charges
do not survive in the nonrelativistic limit: for the extra
supersymmetry parameter $\xi^0$, one can show
\begin{eqnarray}
  \d_\text{extra} Z_\a^I \simeq \xi^0 \bar \Psi_\a^I\ ,
\end{eqnarray}
where $I$ denote the $SO(4)$ gauge indices. Once we choose
the particle modes only, it is highly fluctuating and averages
to zero
\begin{eqnarray}
  \d_\text{extra} z_a^I \simeq e^{2imc^2t} \xi^0 \bar \psi_a^I\ ,
\end{eqnarray}
in the nonrelativistic limit. As mentioned in introduction,
one can in fact keep both particle and anti-particle
in the nonrelativistic limit of the Chern-Simons theory,
compatible with $SO(4)\times SO(4)$. It however leads to
a less interesting nonrelativistic theory with
only sixteen kinematical supercharges, rather
trivial extension of the non-supersymmetric theory
which does not control the dynamics tightly.

Finally, it is known that the Schr\"odinger algebra
can be written in a Virasoro-like form.
\be
{}[L_m, L_n] = (m-n) L_{m+n}\,,
\;\;\;
[L_m, P_r^i ] = \left( \thalf m -r \right) P_{m+r}^i \,,
\;\;\;
[P_r^i , P_s^j ] = (r-s) \d^{ij} \CM \,,
\label{bos-alg-final}
\ee
with the identification
\be
L_{-1} \sim H,\;\; L_0 \sim D,\;\; L_1 \sim K, \;\;
P_{-1/2}^i \sim P^i, \;\; P_{+1/2}^i \sim G^i.
\ee
Moreover, Ref.~\cite{Henkel:1993sg} pointed out that
this algebra admits an infinite dimensional extension
of a Kac-Moody type.
It would be interesting to see whether the
super-Schr\"odinger algebra under discussion also admits such an extension
and if so, how it may help understand the physics.
An infinite dimensional extension of a
non-relativistic conformal algebra
similar to but different from the Schr\"odinger algebra
has been considered recently \cite{Bagchi:2009my}.

%\newpage

\section{Chiral Primary Operators and States}

We are interested in the physical implications of the super Schr\"odinger symmetry on the
nonrelativistic ABJM theory. The theory describes the low energy interaction of particles.
Since the theory is conformal, the most basic physical observables
are the spectrum of conformal operators
and their correlation functions. The characterization of local gauge invariant composite operators would play a crucial role.
%
% it has been an useful exercise to see the representations of operators under the nonrelativistic superconformal, that is super-Schr\"odinger algebra.
The representation theory of (super-)Schr\"odinger algebra
has been discussed in the literature (see, for example, Refs.~\cite{Son:2008ye,Nakayama:2008qm} and references therein). To keep the discussion
self-contained, we reproduce some known results
relevant to our discussion with emphasis on
the new features of the $\CN=6$ algebra.
In this section we put $2m=1$ for notational simplicity.
% of the super Schr\"odinger algebra.

Let us look at the set of all local operators $\CO_a (t,x)$, or quasi-local in our case, and put these operators at the origin $t=x=0$. The representation of the super-Schr\"odinger algebra is realized on
these operators by  the (anti)commutation relation
\be [\CA, \CO_a(0)] = \CA_{ab} \CO_b(0) ,  \ee
for  any generators $\CA$ in this  algebra.  Especially we are interested in the unitary representation, which are realized by the massive particles in the symmetric phase. We mean that
the charges are expressed explicitly as Hermitian operators in the previous section, and so the group realization on the Hilbert space is unitary.

A given operator with definite  scaling dimension $\D_\CO$,   particle number $N_\CO$,  angular momentum $j_\CO$, and $\tilde R$ charge $\tilde r_\CO$ satisfies
\be i[D,\CO ] = \D_\CO \CO   , \;\;\;  [\CN, \CO] = N_\CO \CO , \;\;\; [J,\CO ] = j_\CO \CO , \;\;\;
[\tilde{R},\CO] = \tilde r_\CO \CO .
\ee
%
%Later we will see there is a lower bound on $\D$.
Operators with different quantum numbers can transform onto one another
by the generators of the algebra to form a representation.  As $K$, $G_\pm $ acting on operators would reduce the scaling dimension, one could find the operator of the lowest scaling dimension in a representation.
The {\it conformal primary}  operators
are  defined as the operators which commute with $K$ and $G_i$, that is,
\be [K, \CO(0) ]= 0 , \ [G_\pm, \CO(0) ]=0.   \ee
Each irreducible representation of bosonic Schr\"odinger algebra can be built upon a given primary field.
In our supersymmetric theory, there are also conformal supercharges $S_\pm$ which lower the
scaling dimension. We may wish to define the {\it superconformal primary} operators to be those commuting with
$K,G_\pm, S_\pm$, that is,
\be [K, \CO(0) ]= 0 , \;\;\; [G_\pm, \CO(0) ]=0,\;\;\; [S_\pm, \CO(0)]=0.   \ee
%
%Starting from a superconformal primary operator, we could build up a whole tower of operators of higher scaling dimension by commuting iteratively with other generators.
Similar to   the relativistic superconformal theories, one can define {\it chiral} or {\it antichiral}
primary operators for which some of the dynamical supercharge annihilates:
\be [Q_-,\CO(0)]=0 \ (chiral) , \ \ \ [Q_+, \CO(0)]= 0 \ (antichiral) \ee
However, this definition is somewhat deficient as there are chiral primary fields for which
because the super-Schr\"odinger algebra contains,
unlike its relativistic counterparts,
the kinematic supercharges  $q^{ai}_\pm$
with vanishing scaling dimension.
We will address this issue as well as the problem of BPS-type short multiplets in subsection \ref{ssch-rep}.

\subsection{Operator-state correspondence}

The operator-state map has played a crucial role
in the study of relativistic conformal field theories.
To generalize it to the nonrelativistic case,
let us first recall how it works
in the relativistic case.
In $(d+1)$-dimensions, the Poincar\'e  group $SO(1,d)$
is extended to the $SO(2,d+1)$ conformal group.
The dilatation generator characterizing the spectrum of local operators via
\be
i[D,\CO ] = \D_\CO \CO \,.
\label{dodc}
\ee
is identified with $J_{-1,d+1}$ among the generators $J_{MN}$ ($M,N=-1,0,1,\cdots,d,d+1$) of $SO(2,d+1)$.
The operator-state correspondence asserts that there exists
a one-to-one map,
$\CO \leftrightarrow | \Psi_\CO \rangle$, such that
(\ref{dodc}) translates to
\be
{}\hat{D} | \Psi_\CO \rangle = \D_\CO | \Psi_\CO \rangle .
\label{dodp}
\ee
While (\ref{dodc}) and (\ref{dodp})
share the same eigenvalue $\D_\CO$, $\hat{D}$ is not the same as $D$
but is identified with $J_{0,-1}$.
A canonical way to understand this relation
is the radial quantization; one puts the theory on $\IR \times S^{d}$
with natural action of $SO(2)\times SO(d+1) \subset SO(2,d+1)$,
so that $\hat{D}$ becomes the Hamiltonian.
Alternatively, one can use the fact that
$\hat{D} = J_{0,-1} = \half(P_0 + K_0)$, where $P_\mu$ and $K_\mu$
are translation and special conformal generators.
{}From this point of view, one studies the theory in flat $\IR^{1,d}$
but with the modified Hamiltonian $\hat{D}$ which contains
an explicit space-time dependent term $K_0$ in addition
to the original Hamiltonian.
\footnote{There is yet another approach based on Belavin-Polyakov-Zamolodchikov (BPZ) type conjugation. See \cite{Nakayama:2008qm}
for a discussion in the context of Schr\"odinger symmetry.}

As space and time scale differently in Schr\"odinger symmetric theories,
it is not clear how to generalize the radial quantization.
But, the other approach can be adopted without much difficulty
by using the $SO(2,1)$ subalgebra of Schr\"odinger algebra, as explained recently by Nishida and Son \cite{Nishida:2007pj}
(see also earlier works \cite{deAlfaro:1976je}).
%found a beautiful realization of the operator-state correspondence of the nonrelativistic conformal theory, which is somewhat similar to that radial quantization of the relativistic conformal theory. In relativistic theory, we are putting theory on sphere instead of the
% plane and the Hamiltonian becomes $D$.  In Nishida-Son paper, the original Hamiltonian is modified by adding the harmonic potential given by the special conformal charge $K$.
The additional term $K$ in the modified Hamiltonian
amounts to coupling the theory in an external harmonic potential.
%
%The operator-state correspondence can be used both ways: to find the unitary representation of super-Schrodinger algebra and to study  an energy eigenstates of many particle
%system in a harmonic potential. The primary states would correspond to an special eigenstate of
% the many body system with a harmonic potential.
%
%The scaling dimension of the operator would coincide with the energy eigenvalue of the system. (Recall here we have assumed $2m=1$.)
% Such a operator-state correspondence would lead to other insights about our supersymmetric
% nonrelativistic conformal theories. Especially, the chiral primary operators would lead to the eigenstates of the many body system.

Through the operator-state map, each primary operator  corresponds to an energy eigenstates of a many-body system in a harmonic potential. (Recall here we have assumed $2m=1$.)
The total Hamiltonian is
\be \hat L_0 = H + K . \ee
As $K= \frac14  \int d^2x x_i^2 n(x) $, the potential is confining and preserves the rotational symmetry.
We reorganize the Schr\"odinger algebra by redefining the remaining operators
\be && \hat L_{+1} = \frac12 (H-K-iD), \ \  \hat L_{-1} = \frac12 (H-K + iD) ,  \ \ \hat L_{-1} =
 (\hat L_{+1})^\dagger, \\
&&  \hat P_\pm = P_\pm + iG_\pm, \  \ \hat{G}_\pm = P_\pm -i G_\pm , \ \ \hat G_\pm = (\hat P_\mp)^\dagger . \ee
Note that $L_{\pm 1}$ does not carry any angular momentum.
These generators satisfy the relations
\be && [\hat L_0, \hat L_{\pm 1}] =  \pm 2 \hat L_{\pm 1},
\ [\hat L_{+1}, \hat L_{-1}] = -\hat L_0,  \\
&& [\hat L_0, \hat P_\pm] = \hat P_\pm, \ [\hat L_0 , \hat G_\pm ] = - \hat G_\pm , \
[\hat G_+, \hat P_-] = [\hat G_-,\hat P_+] =   2\CN,  \\
&& [\hat L_{+1} , \hat P_\pm ]= 0 , \ [\hat L_{-1}, \hat P_\pm] = 2\hat G_\pm  ,  \ [\hat L_{-1} , \hat G_\pm ]=0, \ [\hat L_{+1}, \hat G_\pm ] = -2
\hat P_\pm , \ee
and other trivial ones.
For a local operator $\CO$, we can construct  a  state
\be |\Psi_\CO\rangle = e^{-H}\CO (0) |0\rangle .\ee
where $|0\rangle $ is the vacuum of the original Hamiltonian and so is also the vacuum of the new Hamiltonian $\hat L_0$ as our expression of the charges show. This vacuum has no particle and so has zero energy.  For primary operators, we call the corresponding states primary.
 For such primary states, we see
\be && \hat L_0 |\Psi_\CO\rangle = \D_{\CO }|\Psi_\CO\rangle , \   \CN|\Psi_\CO\rangle
=N_{\CO } |\Psi_\CO\rangle,      \  \hat L_{-1} |\Psi_\CO\rangle  = 0 ,   \ \hat G_i |\Psi_\CO\rangle
= 0.  \ee
Thus, primary states are eigenstates of the harmonic hamitonian
$\hat{L}_0$ of energy $\D_\CO$.  Once we find
such primary states, we can find the descendant states, which are also energy eigenstates, by applying the ladder operators $\hat{P}_\pm$ and
$\hat{L}_{+ 1}$. $\hat{P}_\pm$ increases the energy eigenvalues by 1 and
$\hat{L}_{+ 1}$ by 2. These descendant states would correspond to the descendant operators obtained from the primary states by multiple applications of $P_\pm$, $\hat L_{+1}$.
  As the Hamiltonian is invariant under the rotation, we can also choose the primary operator to be an eigenoperator of the angular momentum with value $j_\CO$.   $P_\pm$ changes the angular momentum by $\pm 1$. Thus we can start from the primary state
\be |\Psi_\CO\rangle = | \D_\CO, j_\CO, N_\CO \rangle ,  \ee
and construct all the descendant states
\be \hat{P}_- ^l \hat{P}_+^m \hat{L}_{+1}^n |\Psi_\CO\rangle =
| \D_\CO+ l + m + 2n , J_\CO-l +m , N_\CO \rangle  . \ee
Some of the descendant states could be null.
See \cite{Nakayama:2008qm} for an explicit construction
of the null states.

The unitarity of the Fock space leads to restrictions on the
range of eigenvalues.
\begin{itemize}
\item level-one constraint:
\be ||\hat P_i |\Psi_\CO\rangle ||^2 \ge 0 \Longrightarrow   N_\CO \ge 0.  \ee
\item level-two constraint:
\be ||(2\CM \hat L_{+1}   - \hat P_i \hat P_i)|\Psi_\CO\rangle ||^2 \ge 0 \Longrightarrow\  \D_\CO\ge 1 \ \ \text{if} \ \   N_\CO\neq 0 . \ee
($\D_\CO \ge d/2$ in $d$-space dimension.) The bound is saturated when
\be \hat L_{+1} |\Psi_\CO\rangle = \frac{1}{2\CM} \hat P_i^2 |\Psi_\CO \rangle . \ee
The dimension $1$ operator satisfies the free Schr\"odinger equation.
\end{itemize}
The restriction $\D_\CO\ge d/2$ could be understood as the zero-point energy of a particle in a harmonic oscillator in $d$ dimensions \cite{Nishida:2007pj}.

For the relativistic examples, the radial quantization of the conformal
field theory on the $\mathbb{R}\times S^{d}$ can be naturally
understood in the context of AdS$_{d+2}$. It will be very
interesting to understand the physical origins of this
nonrelativistic system in the dual gravity picture.

\subsection{Representation of super-Schr\"odinger algebra \label{ssch-rep}}

We first characterize a local operator $\CO$ by its scaling dimension $\D_\CO$, spin $j_\CO$,
particle number $N_\CO$, and $\tilde{R}$ charge $r_\CO$, and irreducible representations $(r_1,r_2)$ ($r_i \in \half \IZ$) of the $SU(2)\times SU(2)$ $R$-symmetry.
To take advantage of the operator-state map, let us reorganize
the dynamical supercharges as
\be && \hat Q_+   = Q_+  -i S_+   , \ \hat S_-
= Q_-   + i S_-   = (\hat{Q}_+)^\dagger ,  \\
&& \hat Q_- = Q_-  -i S_-, \ \hat S_+  = Q_+ + i S_+ = (\hat Q_-)^\dagger . \ee
They satisfy the following relations:
\be && \{ \hat Q_-, \hat S_+\} = \hat L_0 - (J - \frac32 \tilde R) ,   \\
&& \{ \hat Q_+, \hat S_-\} = \hat L_0 +(J - \frac32 \tilde  R) ,   \\
&& \{ \hat Q_-, \hat Q_+    \}  = 2 \hat L_{+1}, \ \{ \hat S_-,\hat S_+\} = 2\hat L_{-1}.
\ee
In addition,  we have the relations
\be  &    [\hat L_0, \hat Q_\pm] =\hat Q_\pm, & [\hat L_{+1}, \hat Q_\pm] = 0
, \ \ \ \  \ \  [\hat L_{-1}, \hat Q_\pm] = \hat S_\pm ,\\
&   [\hat L_0 , \hat S_\pm] = -\hat S_\pm, & [\hat L_{+1}, \hat S_\pm]
= -\hat Q_\pm, \ \,  [\hat L_{-1}, \hat S_\pm]=0 .
\ee
The remaining nonzero commutators are
 \be [\hat P_\pm, \hat S_\mp ]  = -2q_\pm  , \ [\hat G_\pm, \hat Q_\mp] =  +2q_\pm ,  \
 \{ q_\pm , \hat Q_\pm\} =  \hat P_\pm, \ \{ q_\pm , \hat S_\pm\} = \hat G_\pm . \ee

\begin{figure}[t]
\begin{center}
\includegraphics[width=14cm]{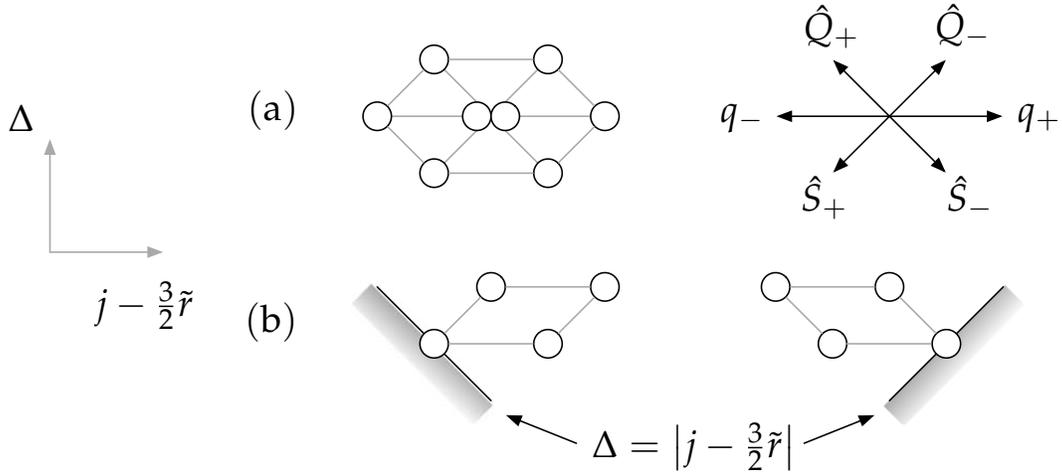}
\caption{Multiplet structure of the super-Schr\"odinger algebra.
(a) A long multiplet consists of eight conformal primary operators.
(b) A short (chiral or anti-chiral) multiplet
consists of four conformal primary operators.
} \label{short-long}
\end{center}
\end{figure}

An irreducible representation (irrep) of the super-Schr\"odinger
algebra ${\bf sch_2^{(1)}}$  should consist of
several irreps of the bosonic Schr\"odinger algebra ${\bf sch_2^{(0)}}$.
As the dynamical part of the superalgebra
contains three pairs of fermionic oscillators
($Q_\pm, S_{\pm}, q_{\pm}$), we expect generically
eight irreps of the bosonic algebra to form a multiplet.
As mentioned earlier, the irreps of the bosonic algebra are specified
by their conformal primary states,
so in order to specify a super-multiplet,
it suffices to show how the eight primary states
get mapped to each other by the super-charges.

The structure of a long multiplet of ${\bf sch_2^{(1)}}$  is depicted in
Figure~\ref{short-long}(a).
We begin with the primary state $| 1 \rangle $ with the lowest
value of $\Delta_\CO$. By assumption it satisfies
$\hat{L}_{- 1} | 1 \rangle =\hat{G}_{\pm}| 1 \rangle = \hat{S}_{\pm} | 1 \rangle =0$.
We further assume that $q_- |1 \rangle = 0$.
When $N_\CO \neq 0$, we can rescale the generators
such that
\be
\{ q_+, q_- \} = 1,
\qquad {}[\hat{G}_\pm, \hat{P}_{\mp} ] = 2,
\ee
with other commutation relations unchanged.
Then $|1\rangle$ is naturally paired with another primary operator
$| 2 \rangle = q_+ | 1 \rangle$ with the same
$\Delta_\CO$. Thus a generic (long) multiplet
is specified by a pair of superconformal primary
($\hat{S}_\pm |\psi\rangle = 0$) states.
The remaining six primary states in the multiplet can be written
explicitly as follows:
\be
| 3 \rangle = - \hat Q_+ | 1 \rangle \,,
&&
| 4 \rangle = \hat Q_+ | 2 \rangle - \hat P_+ | 1 \rangle \,,
\nn \\
| 5 \rangle = \hat Q_- | 1 \rangle - \hat P_ -| 2 \rangle \,,
&&
| 6 \rangle = -\hat Q_- | 2 \rangle \,,
\nn \\
| 7 \rangle = \hat Q_- | 3 \rangle - \hat Q_+ | 5 \rangle -\hat P_- | 4 \rangle \,,
&&
| 8 \rangle = \hat Q_+ | 6 \rangle - \hat Q_- | 4 \rangle - \hat P_+ | 5 \rangle \,.
\label{long-multi}\ee
It is easy to show that all of these are primary and have finite norm.
They also satisfy $q_+ | 2k-1 \rangle = |2k \rangle$.
Note that, in general, the action of $\hat{Q}_\pm$ or $\hat{S}_\pm$
on a primary state in a multiplet yields
a linear combination of other primary states
as well as some descendants.
%(Similar things happen in relativistic superconformal algebras).

This structure should be augmented by the fact that
our $\CN=6$ algebra ${\bf sch_{2}^{(2)}}$  includes additional supercharges
$q^{ai}_\pm$ and the $SU(2)\times SU(2)$ $R$-symmetry generators.
However, since these generators commute with the dynamical
generators, we can simply take the tensor product
of the irreps on each side. We will come back to
the representation theory of the ($q^{ai}_\pm, R_1, R_2$)
subalgebra shortly.

Next, let us consider unitarity constraints of the super-Schr\"odinger algebra. Note that for any  state   $|\Psi_\CO\rangle =e^{-H}\CO|0\rangle $ of charges $\D_\CO, j_\CO, N_\CO, r_\CO$,
\be || \hat Q_\pm |\Psi_\CO\rangle ||^2  + || \hat S_\mp|\Psi_\CO\rangle ||^2 = \langle \Psi_\CO|\{\hat S_\mp, \hat Q_\pm \} |\Psi_\CO\rangle \ge 0 , \ee
which leads to an inequality
\be \D_\CO \ge \left| j_\CO - \frac32 r_\CO \right|.  \label{dbound}  \ee
The bound is saturated by three types of short multiplets.
%besides the vacuum state.

\noindent{I.}  {\it chiral}  operators or states for which
\be  \D_\CO = j_\CO -\frac32  r_\CO \Longrightarrow [Q_-,\CO]= [S_+,\CO]=0
 \ {\rm and}  \  \hat Q_-|\Psi_\CO\rangle = \hat S_+|\Psi_\CO\rangle = 0   , \ee
\noindent{II.} {\it anti-chiral}  operators or states for which
\be  \D_\CO = - j_\CO + \frac32  r_\CO \Longrightarrow [Q_+,\CO]= [S_-,\CO]=0  \ {\rm and}  \  \hat Q_+|\Psi_\CO\rangle = \hat S_-|\Psi_\CO\rangle = 0   , \ee
\noindent{III.} {\it vacuum} operators or states for which
\be \D_\CO=   \ j_\CO- \frac32 r_\CO =0  \Longrightarrow [Q_\pm, \CO]= [S_\pm,\CO]=0  \ {\rm and}  \  \hat Q_\pm |\Psi_\CO\rangle = \hat{S}_\pm |\Psi_\CO\rangle = 0 . \ee
The identity operator (vacuum state) with zero spin and zero $R$-charge
appears to be the only vacuum operator in our theory.

One sees chiral primary operators (states)  and anti-chiral primary operators saturate the unitarity bound. There are also the {\it chiral descendant}  operators that are chiral ({\it i.e.} saturate the bound) but not superconformal primary.  The chiral  descendants can be obtained from the chiral  primaries  by multiple applications of $\hat P_+$, which increases not only $\D$ but also angular momentum $j$.
The chiral primary and descendant states
contribute to the superconformal index we will discuss in section \ref{indexx}.

One can build a short representation of the super-Schr\"odinger algebra ${\bf sch_{2}^{(1)}}$ starting from a given chiral primary state. Each short multiplet contains four primary states, as shown in~\ref{short-long}.
If the theory admits a continuous deformation,
the dimension $\D_\CO$ of a long multiplet may  be lowered until it hits the unitarity bound. Then the long multiplet can split
into two short multiplets. The pattern of the splitting
should can be seen clearly in Figure~\ref{short-long}, (a) and (b).

Finally, let us turn to the representation of the superalgebra ${\bf sch_2^{(2)}}$ whose kinematical supercharges satisfy the algebra
\be
\{ q_-^{\ ai}, q_{+ bj}\}  =  \frac12 \delta^a_b \d^i_j \CN
- \delta^a_b R^i_{\ j} +\d^i_j R^a_{\ b}     .
\label{k4susy2}
\ee
where $R^a_{\ b}, R^i_{\ j}$ denote generators of R-symmetry group $SU(2)\times SU(2)$.
Its representation theory has been studied in \cite{Beisert:2007pu}.
For self-containedness, we review it in our context. 

In a given irreducible representation of algebra ${\bf sch_2^{(1)}}$,
let us choose any primary state $|\Psi\rangle$
of number charge $N$ and possible lowest angular momentum 
$j_0$ which transforms as an irreducible representation 
$(r_1,r_2)$ of two $SU(2)$'s. The operators $q^{ai}_- $ and $q_{+ai}$  
lowers and raises the angular momentum $j$ by 1/2, respectively, 
but does not change the eigenvalues of $(J -\frac{3}{2}\tilde{R})$.
As $|\Psi\rangle=|N,j_0,r_1,r_2\rangle $ has the lowest angular momentum,
\begin{eqnarray}  
  q_-^{\ ai} |N,j_0,r_1,r_2\rangle=0 , \ \text{for all} \ \ a,i \ . 
\end{eqnarray}
For the unitarity, one has to demands that the matrix below 
\begin{eqnarray} 
  \langle\Psi' | q_-^{\ ai} q_{+ bj} | \Psi\rangle =
  \frac12 \d^a_b \d^i_j \langle \Psi' | \CN |  \Psi\rangle   - \d^a_b \langle \Psi'|  R^i_{\ j} |\Psi\rangle  + \d^i_j \langle \Psi'| R^a_{\ b} |\Psi \rangle  .
\end{eqnarray}
has only non negative eigenvalues. After a suitable diagonalization, one can 
obtain the lower bound on number charge $\CN$
\begin{eqnarray}
  \CN \geq2~\text{max}\left[
  \Big( c_2\big(r_1\big)+ c_2\big(\frac12\big) -c_2\big(r'_1\big) \Big)
  + \Big( c_2\big(r'_2\big) - c_2\big(r_2\big) - c_2\big(\frac12\big) \Big) \Big|
  \ r_i \otimes \frac12 =\oplus r'_i \right]\ ,
  \end{eqnarray}
where $c_2(r)$ is the quadratic Casimir of $SU(2)$ representation
$r$, normalized such that $c_2(r)=r(r+1)$. 
The above bound is saturated when $c_2(r'_1)$
takes the smallest value while $c_2(r'_2)$ takes 
the largest one. The unitary restriction on 
number charge $\CN$ is therefore given by
\begin{eqnarray}\label{nbound}
  \left\{
  \begin{array}{lc}
  \CN \geq 2\big( r_1 + r_2 +1 \big) & (r_1 \neq 0) \\
  \CN \geq 2 r_2 & (r_1 = 0) 
  \end{array}
  \right. \ .
\end{eqnarray}

Applying the  operators $q_{+ ai}$ changes the representations $r_1, r_2$ by $\pm 1/2$, and so generates
$2^4(2r_1+1)(2r_2+1) $ primary states for the same $\D$ and
$(J-\frac{3}{2}\tilde{R})$ eigenvalues:
\begin{itemize}

\item
$j=j_0$ or $j_0+2$ : $(r_1,r_2)$ .

\item
$j=j_0+1/2$ or $j_0+3/2$ : $(r_1\pm1/2,r_2\pm1/2)$   .

\item
$j=j_0+1$ : $(r_1, r_2\pm 1)\oplus (r_1\pm 1, r_2) \oplus 2(r_1,r_2)$ .

\end{itemize}
Of course, if some of  $r_1$ or $r_2$ is less than or equal to $1/2$, the number of the
derived representations would be reduced as some of the derived irreducible representations does not exist.
When the  starting state is close to saturate the number unitary bound (\ref{nbound}),
then again some of the number of the derived states would become null
and the number of derived states would be reduced.

We say the superconformal primary operators is {\it bps}  operators or states
if the above number unitary bound (\ref{nbound}) is saturated.
Those states are now annihilated by some of creation operators $q_{+ai}$ 
in such a manner as to give us the smallest $c_2(r_1')$ and largest $c_2(r_2')$.
The resulting multiplet contains a total of $2^3(4r_1r_2 + r_1+3r_2+1)$ states,
which can be decomposed into the following bosonic irreps:
\begin{itemize}

\item
$j=j_0$   : $(r_1,r_2)$.

\item
$j=j_0+1/2$   : $(r_1 + 1/2,r_2 + 1/2) \oplus (r_1 - 1/2,r_2 - 1/2) \oplus (r_1 + 1/2,r_2 - 1/2) $.

\item
$j=j_0+1$ : $(r_1+1 , r_2)\oplus (r_1, r_2-1) \oplus (r_1,r_2)$.

\item
$j=j_0+3/2$:  $(r_1+1/2, r_2-1/2)$.

\end{itemize}
Again the above derivation would be reduced if $r_1$ or $r_2$ is less than or equal to 1/2.

For a fixed value of $(r_1, r_2)$, a long multiplet can split into
two short multiplets if the value of $N_\CO$ is lowered to saturate
the {\it bps} bound (\ref{nbound}). If we denote the multiplets by
$[N_\CO,r_1,r_2]_{\rm long/short}$, the splitting rule is
($N_\CO=2(r_1+r_2+1)$)
\be
[N_\CO,r_1,r_2]_{\rm long} &\goto& [N_\CO, r_1,r_2]_{\rm short} + [N_\CO, r_1-1/2,r_2+1/2]_{\rm short}  \,,
\nn \\
2^4(2r_1+1)(2r_2+1) &=& 2^3(4r_1r_2+r_1+3r_2+1)+ 2^3(4r_1r_2+3r_1+r_2+1) \,.
\ee

\subsection{Elementary fields and states}

For the $U(1)\times U(1)$ theory, we can introduce the magnetic  flux operator and multiply
it to the fields $z_\a, \psi^\a$ to make them gauge invariant.
The flux operator carries fractional magnetic flux, making the physical operator to be quasi-local.
However these charge-flux composite operators are not of anyonic ones with fractional spin
and statistics, as the charge and the flux do not know each other.
\begin{table}
\caption{ Charges for the physical fields in $U(1)\times U(1)$ theory }
\begin{center}
\begin{tabular}{|c|cccc|c|c|}
 \hline
 & $\D $ & $ J $ & $2R $ &  $  2\Sigma $& $ \tilde R=(4R+2\Sigma)/3 $ & $J-\frac32 \tilde{R} $  \\
 \hline
 $\bar z^a $     & $ 1$   & $ 0$       & $+1$   & $   0$  & $+2/3$  & $-1$ \\
  $\bar z^i $      & $1 $  & $ 0 $      & $-1 $  & $ 0$    & $-2/3$  & $+1$ \\
  $\bar \psi_a $ & $ 1$  & $+1/2 $  &  $ -1$  & $ +1$ & $-1/3$ &  $+1$ \\
  $\bar \psi_i $  & $ 1$  & $-1/2  $   & $+ 1$     & $-1$   & $+1/3$ & $-1$ \\
\hline
\end{tabular}
\end{center}
\label{gcharges2}
\end{table}
\begin{table}
\caption{ Physical fields under the supersymmetry }
\begin{center}
\begin{tabular}{|c|cccc|}
 \hline
 & $Q_+ $ & $Q_- $ & $q_+$ &  $  q_- $   \\
 \hline
 $\bar z^a $  & $ 0 $
  & $    -i\e^{ab}D_-\bar\psi_b $
   & $ -\e^{ab}\bar\psi_b$  & $   0$    \\
  $\bar z^i $      & $ i\e^{ij}D_+\bar \psi_j $  & $ 0 $  & $ 0$  & $ \e^{ij}\bar \psi_j$
     \\
  $\bar \psi_a $ & $ -i \e_{ab}D_+\bar z^b$  & $0 $  &  $ 0$  & $ -\e_{ab}\bar z^b $   \\
  $\bar \psi_i $  & $ 0 $  & $ i\e_{ij} D_- \bar z^j  $   & $ \e_{ij}\bar z^j $     & $0$     \\
\hline
\end{tabular}
\end{center}
\label{gcharges3}
\end{table}
For convenience, we summarize the relevant charges of physical fields in Table \ref{gcharges2}
and the kinematical/dynamical supercharges by which physical fields are annihilated
in Table \ref{gcharges3}.

One can analyze what representation the creation operators in the $U(1)\times U(1)$ theory form
under the super Schr\"odinger algebra ${\bf sch_2^{(2)}}$. One can see that
elementary physical fields at the origin saturate the scaling-dimension bound (\ref{dbound})
and  are split  to chiral and anti-chiral primary
operators. Also they saturate the number bound (\ref{nbound}) and so are  bps operators:
\be \begin{array}{lll}
   \bar z^i(0), \bar\psi_{+a}(0):   &  ( 1/2,  0)  &  \text{bps chiral  primary}   \\
  \bar z^a(0) , \bar\psi_{-i}(0) : &  ( 0, 1/2) &  \text{bps anti-chiral primary}
\end{array}\label{u1case} \ee
In the $\CN=4$ language,  the first set of the fields $\bar z^i, \bar\psi_a$ came from
the twisted hyper-multiplet and the second set of the field $\bar z^a, \bar\psi_i$ came
from the hyper-multiplet.
The products of chiral fields $\bar z^i$ and $\bar\psi_a$ will remain   chiral primary.
In addition, they would saturate   the  number bound (\ref{nbound})
if one symmetrizes  the $SU(2)\times SU(2)$ indices.

The hamiltonian $\hat L_0$ has a harmonic potential.
Let us imagine a bunch of particles rotating around the origin in circular orbits.
The radial kinetic energy can be ignored and the total energy would be roughly
\be \D \sim \frac{mw^2}{2} r^2 + \frac{J^2}{2mr^2} \,,\ee
which is minimized when $J \sim mwr^2$. The energy becomes
\be \D\sim wJ \,. \ee
Thus,   our chiral states in large $\D$ limit  may be represented by such configurations.
Indeed, for $U(1)\times U(1)$ case, the chiral operator  $\partial_+^n \bar z^i$ with
large $n$ may represent such circular motion.

For a larger gauge group, we argued in Sec. 2 for the existence of  the nonabelian flux-charge comoposte
creation operators,
\be \bar {\cal Z}^i,\ \bar\varPsi_{+a}, \ \  \bar {\cal Z}^a, \   \bar\varPsi_{-i} .  \ee
They are invariant under the local gauge transformation but transform
as $N\times \bar N$ under the global part of the $U(N)\times U(N)$
gauge symmetry. They would discribe the creation operators for
single massive particle. Thus their spin and R-quantum number  would be more and
less identical to that of the abelian case. Thus these operators would
be also bps (anti-)chiral operators. In addition, as they carry the
fractional nonabelian magnetic flux, these composite operators may be
quasi-local in the sense their statitics may be fractional. It would
be interesting to construch such operators explicitly, at least
perturbatively in the weak coupling or large $k$ regime, and to
explore their two point functions.

In addition, the quantum partners of the classical
BPS configuration of our Lagrangian would play a special role. The
detail study  of the classical BPS configurations need? some attention. 
  In our theory there exist di-baryonic operators, say,
made of only the scalar fields $z^{\ A}_{a\ M}(0)$ without any flux attached,
which would be invariant under the local gauge transformation, and
also singlet under the global part of the gauge group.  These
di-baryonic operators would remain chiral.

\subsection{Comments on the superconformal index \label{indexx}}

Using the operator-state correspondence, one can
naturally define a nonrelativistic
superconformal Witten index \cite{Nakayama:2008qm}
in analogy with the relativistic counterpart \cite{Kinney:2005ej}, which counts the chiral states that are annihilated by $\hat Q_-, \hat S_+$.\footnote{For the relativistic ABJM model and its orbifolds,
the index has been studied in \cite{Bhattacharya:2008bja,Dolan:2008vc,Choi:2008za}.}
They are made of chiral primary and descendant states.  For general states,
there are two bounds    (\ref{dbound}) and (\ref{nbound}).
The first bound is saturated by the chiral states.
In addition, there are many ways to discern these chiral states by measuring their
additional charges. The examples are
\be \D + 2J , \  N- 2(r_1+r_2) , r_1-r_2, \ m_1, \ m_2 , \ee
which commute  with $Q_-, S_+$. The index we want to compute is therefore given by
\be \CI(x,y_1,y_2) = \lim_{\b \rightarrow \infty} {\rm Tr}
\Big[ (-1)^F e^{-\beta (\D - (j-\frac{3}{2} r) )} e^{ \mu (N - 2(r_1+r_2)) }
x^{\D+2j} y_1^{m_1} y_2^{m_2} y_3^{r_1-r_2}\Big]  \ee
In the $\mu\rightarrow -\infty$ limit, the above index counts only $bps$ states.

The superconformal index of a relativistic theory
in the limit of vanishing coupling can be computed in two steps \cite{Kinney:2005ej}.
One begins by computing the index for chiral `letters', namely,
elementary fields and derivatives without worrying about gauge invariance. The result is then inserted into a matrix integral
which efficiently picks out the gauge invariant combinations
among products of elementary fields and derivatives.

The chiral ($\D-J+\frac32 \tilde R \doteq 0$) `letters' of the nonrelativistic ABJM model are just $\bar z^i, \bar \psi_a $ and
covariant derivatives $D_+$. Note that the field strength is not chiral. The fields $z_a, \psi^i$ also have the correct quantum numbers
to become chiral letters,
but since they are {\em annihilation} operators,
they could not contribute to creating an energy eigenstate
of the many body system in a harmonic potential.
If we stick to $\bar{z}^i$ and $\bar{\psi}_a$ only,
we could not form any gauge-invariant `mesonic' operators
(trace of products of bi-fundamental fields).
The `di-baryonic' operators involving determinants
of the $U(N)\times U(N)$ gauge groups are not affected. It is possible to count the
mesonic operators in the $N=0$ section as done in Ref.~\cite{Nakayama:2008qm},
but its physical interpretation seems unclear to us.

We think that the  flux-charge composite operators $\bar{\CZ}^i, \bar\varPsi_a$
remain chiral even in the nonabelian case as in the abelian case. They are
invariant under the local gauge transformation, but become bi-fundamental under
the global part of the gauge group. These operators would create the chiral states
and would contribute to the chiral index. It would be very interesting to find out whether
this is the case in the large $k$ limit where the Gauss law becomes simpler.

\vskip 2cm

\subsection*{Acknowledgment}

It is our pleasure to thank Eoin \'O Colg\'ain, Hossein Yavartanoo, Rajesh Gopakumar and Seok Kim for
helpful discussions. We also thank to Yu Nakayama for discussions on subtleties in
superconformal index. KML is supported in part by the
KOSEF SRC Program through CQUeST Sogang University and KRF-2007-C008 program.
Sm.L. is supported in part by the KOSEF Grant R01-2006-000-10965-0 and the
Korea Research Foundation Grant KRF-2007-331-C00073.

\vskip 2cm

%\newpage


\begin{thebibliography}{99}
\parskip 0.0cm


%%%%%%%%%%%%%%%%%%%%%%%%%%%%%%%%%%%%%%%%%%%%%%%%%%%%%%%%%%%%%%%%%%
%%%
%%% Introduction begins here
%%%

%%%%%%%%%%%%%%%%%%%%%%%%%%%%%%%%%%%%%%%%%%%
%%% Essentials in NR gravity background %%%
%%%%%%%%%%%%%%%%%%%%%%%%%%%%%%%%%%%%%%%%%%%

\bibitem{Son:2008ye}
  D.T.~Son,
  ``{\it Toward an AdS/cold atoms correspondence: a geometric realization of the
  Schroedinger symmetry},''
  Phys.\ Rev.\  D {\bf 78}, 046003 (2008)
  [arXiv:0804.3972 [hep-th]].

\bibitem{Balasubramanian:2008dm}
  K.~Balasubramanian and J.~McGreevy,
  ``{\it Gravity duals for non-relativistic CFTs},''
  Phys.\ Rev.\ Lett.\  {\bf 101}, 061601 (2008)
  [arXiv:0804.4053 [hep-th]].

\bibitem{Herzog:2008wg}
  C.P.~Herzog, M.~Rangamani and S.F.~Ross,
  ``{\it Heating up Galilean holography},''
  JHEP {\bf 0811}, 080 (2008)
  [arXiv:0807.1099 [hep-th]].

\bibitem{Maldacena:2008wh}
  J.~Maldacena, D.~Martelli and Y.~Tachikawa,
  ``{\it Comments on string theory backgrounds with non-relativistic conformal
  symmetry},''
  JHEP {\bf 0810}, 072 (2008)
  [arXiv:0807.1100 [hep-th]].

\bibitem{Adams:2008wt}
  A.~Adams, K.~Balasubramanian and J.~McGreevy,
  ``{\it Hot Spacetimes for Cold Atoms},''
  JHEP {\bf 0811}, 059 (2008)
  [arXiv:0807.1111 [hep-th]].

\bibitem{Hartnoll:2008rs}
  S.~A.~Hartnoll and K.~Yoshida,
  ``{\it Families of IIB duals for nonrelativistic CFTs},''
  JHEP {\bf 0812}, 071 (2008)
  [arXiv:0810.0298 [hep-th]].

\bibitem{Donos:2009en}
  A.~Donos and J.P.~Gauntlett,
  ``{\it Supersymmetric solutions for non-relativistic holography},''
  arXiv:0901.0818 [hep-th].

%%%%%%%%%%%%%%%%%%%%%%%%%%%%%%%%%%%%%%%%%%
%%% Early history of Sch. symmetry     %%%
%%%%%%%%%%%%%%%%%%%%%%%%%%%%%%%%%%%%%%%%%%

\bibitem{Hagen:1972pd}
  C.R.~Hagen,
 ``{\it Scale and conformal transformations in galilean-covariant field theory},''
   Phys.\ Rev.\  D {\bf 5}, 377 (1972).

\bibitem{Niederer:1972zz}
  U.~Niederer,
  ``{\it The maximal kinematical invariance group of the free Schrodinger
  equation},''
  Helv.\ Phys.\ Acta {\bf 45}, 802 (1972).


%%%%%%%%%%%%%%%%%%%%%%%%%%%%%%%%%%%%%%%%%%
%%% Condensed matter system with Sch.  %%%
%%%%%%%%%%%%%%%%%%%%%%%%%%%%%%%%%%%%%%%%%%

\bibitem{Nishida:2007pj}
  Y.~Nishida and D.T.~Son,
  ``{\it Nonrelativistic conformal field theories},''
  Phys.\ Rev.\  D {\bf 76}, 086004 (2007)
  [arXiv:0706.3746 [hep-th]].




%%%%%%%%%%%%%%%%%%%%%%%%%%%%%%%%%%%%%%%%%%%%%
%%% ABJM & BLG and their mass-deformation %%%
%%%%%%%%%%%%%%%%%%%%%%%%%%%%%%%%%%%%%%%%%%%%%

\bibitem{Aharony:2008ug}
  O.~Aharony, O.~Bergman, D. L.~Jafferis and J.~Maldacena,
  ``{\it ${\cal N}=6$ superconformal Chern-Simons-matter theories,
  M2-branes and their gravity duals},''
  arXiv:0806.1218 [hep-th].
  %%CITATION = ARXIV:0806.1218;%%

\bibitem{Hosomichi:2008jd}
  K.~Hosomichi, K.M.~Lee, S.~Lee, S.~Lee and J.~Park,
  ``{\it ${\cal N}=4$ superconformal Chern-Simons theories
  with hyper and twisted hyper multiplets},''
  JHEP {\bf 0807}, 091 (2008)
  arXiv:0805.3662 [hep-th].
  %%CITATION = JHEPA,0807,091;%%

\bibitem{Hosomichi:2008jb}
  K.~Hosomichi, K.M.~Lee, S.~Lee, S.~Lee and J.~Park,
  ``{\it ${\cal N}=5,6$ superconformal Chern-Simons theories
  and M2-branes on orbifolds},''
  arXiv:0806.4977 [hep-th].
  %%CITATION = ARXIV:0806.4977;%%

\bibitem{Gomis:2008vc}
  J.~Gomis, D.~Rodriguez-Gomez, M.~Van Raamsdonk and H.~Verlinde,
  ``{\it A massive study of M2-brane proposals},''
  arXiv:0807.1074 [hep-th].
  %%CITATION = ARXIV:0807.1074;%%

\bibitem{Bagger:2006sk}
  J. Bagger and N. Lambert,
  ``{\it Modeling multiple M2's},''
  Phys.\ Rev.\  D {\bf 75}, 045020 (2007)
  [arXiv:hep-th/0611108].
  %%CITATION = PHRVA,D75,045020;%%

\bibitem{Bagger:2007jr}
  J. Bagger and N. Lambert,
  ``{\it Gauge symmetry and supersymmetry of multiple M2-branes},''
  Phys.\ Rev.\  D {\bf 77}, 065008 (2008)
  arXiv:0711.0955 [hep-th].
  %%CITATION = PHRVA,D77,065008;%%

\bibitem{Bagger:2007vi}
  J. Bagger and N. Lambert,
  ``{\it Comments On multiple M2-branes},''
  JHEP {\bf 0802}, 105 (2008)
  arXiv:0712.3738 [hep-th].
  %%CITATION = JHEPA,0802,105;%%

\bibitem{Gustavsson:2007vu}
  A. Gustavsson,
  ``{\it Algebraic structures on parallel M2-branes},''
  arXiv:0709.1260 [hep-th].
  %%CITATION = ARXIV:0709.1260;%%

\bibitem{Gustavsson:2008dy}
  A.~Gustavsson,
  ``{\it Selfdual strings and loop space Nahm equations},''
  JHEP {\bf 0804}, 083 (2008)
  arXiv:0802.3456 [hep-th].

\bibitem{Gomis:2008cv}
  J.~Gomis, A.~J.~Salim and F.~Passerini,
  ``{\it Matrix theory of type IIB plane wave from membranes},''
  JHEP {\bf 0808}, 002 (2008)
  [arXiv:0804.2186 [hep-th]].
  %%CITATION = JHEPA,0808,002;%%

\bibitem{Hosomichi:2008qk}
  K.~Hosomichi, K.~M.~Lee and S.~Lee,
  ``{\it Mass-deformed Bagger-Lambert theory and its BPS objects},''
  arXiv:0804.2519 [hep-th].
  %%CITATION = ARXIV:0804.2519;%%


%%%%%%%%%%%%%%%%%%%%%%%%%%%%%%%%%%%%%%%%%%%
%%% Geometry for the mass-deformed M2's %%%
%%%%%%%%%%%%%%%%%%%%%%%%%%%%%%%%%%%%%%%%%%%

\bibitem{Bena:2000zb}
  I.~Bena,
  ``{\it The M-theory dual of a 3 dimensional theory with reduced supersymmetry},''
  Phys.\ Rev.\  D {\bf 62}, 126006 (2000)
  [arXiv:hep-th/0004142].

\bibitem{Bena:2004jw}
  I.~Bena and N.~P.~Warner,
  ``{\it A harmonic family of dielectric flow solutions with maximal
  supersymmetry},''
  JHEP {\bf 0412}, 021 (2004)
  [arXiv:hep-th/0406145].




%%%%%%%%%%%%%%%%%%%%%%%%%%%%%%%%%%%%%%%%%
%%% Field theory realization of SSCh. %%%
%%%%%%%%%%%%%%%%%%%%%%%%%%%%%%%%%%%%%%%%%

\bibitem{Jackiw:1990mb}
  R.~Jackiw and S.Y.~Pi,
  ``{\it Classical and quantal nonrelativistic Chern-Simons theory},''
  Phys.\ Rev.\  D {\bf 42}, 3500 (1990)
  [Erratum-ibid.\  D {\bf 48}, 3929 (1993)].

\bibitem{Leblanc:1992wu}
  M.~Leblanc, G.~Lozano and H.~Min,
  ``{\it Extended superconformal Galilean symmetry in Chern-Simons matter systems},''
  Annals Phys.\  {\bf 219}, 328 (1992)
  [arXiv:hep-th/9206039].

\bibitem{Sakaguchi:2008rx}
  M.~Sakaguchi and K.~Yoshida,
  ``{\it Super Schrodinger in Super Conformal},''arXiv:0805.
  2661 [hep-th].

\bibitem{Sakaguchi:2008ku}
  M.~Sakaguchi and K.~Yoshida,
  ``{\it More super Schrodinger algebras from $psu(2,2|4)$},''
  JHEP {\bf 0808}, 049 (2008)
  [arXiv:0806.3612 [hep-th]].

\bibitem{Sakaguchi:2008zz}
  M.~Sakaguchi and K.~Yoshida,
  ``{\it Super Schroedinger algebra in AdS/CFT},''
  J.\ Math.\ Phys.\  {\bf 49} (2008) 102302.



\bibitem{Nakayama:2008qm}
  Y.~Nakayama,
  {``\it Index for non-relativistic superconformal field theories},''
  JHEP {\bf 0810}, 083 (2008)
  [arXiv:0807.3344 [hep-th]].

\bibitem{Nakayama:2008qz}
  Y.~Nakayama, S.~Ryu, M.~Sakaguchi and K.~Yoshida,
  ``{\it A family of super Schrodinger invariant Chern-Simons matter systems},''
  JHEP {\bf 0901}, 006 (2009)
  [arXiv:0811.2461 [hep-th]].


%\cite{Dunne:1990qe}
\bibitem{Dunne:1990qe}
  G.~V.~Dunne, R.~Jackiw, S.~Y.~Pi and C.~A.~Trugenberger,
    ``{\it Selfdual Chern-Simons solitons and two-dimensional nonlinear equations},''
  Phys.\ Rev.\  D {\bf 43} (1991) 1332
  [Erratum-ibid.\  D {\bf 45} (1992) 3012].
  %%CITATION = PHRVA,D43,1332;%%




%%%%%%%%%%%%%%%%%%%%%%%%%%%%%%%%%%%%%%%%
%%% Cocyles in Chern-Simons theories %%%
%%%%%%%%%%%%%%%%%%%%%%%%%%%%%%%%%%%%%%%%

%\cite{Dunne:1989cz}
\bibitem{Dunne:1989cz}
  G.~V.~Dunne, R.~Jackiw and C.~A.~Trugenberger,
  ``{\it Chern-Simons theory in the Schr\"{o}dinger representation},''
  Annals Phys.\  {\bf 194} (1989) 197.
  %%CITATION = APNYA,194,197;%%

%\cite{Elitzur:1989nr}
\bibitem{Elitzur:1989nr}
  S.~Elitzur, G. W.~Moore, A.~Schwimmer and N.~Seiberg,
  ``{\it Remarks on the canonical quantization of
  the Chern-Simons-Witten theory},''
  Nucl.\ Phys.\  B {\bf 326} (1989) 108.
  %%CITATION = NUPHA,B326,108;%%



%%%%%%%%%%%%%%%%%%%%%%%%%%%%%%%%%%%%%%%%%
%%% N=4 kin. superalgebra and SU(2|2) %%%
%%%%%%%%%%%%%%%%%%%%%%%%%%%%%%%%%%%%%%%%%

\bibitem{Lin:2005nh}
  H.~Lin and J.M.~Maldacena,
  ``{\it Fivebranes from gauge theory},''
  Phys.\ Rev.\  D {\bf 74}, 084014 (2006)
  [arXiv:hep-th/0509235].

\bibitem{Agarwal:2008pu}
  A.~Agarwal, N.~Beisert and T.~McLoughlin,
  ``{\it Scattering in mass-deformed $\CN\geq4$ Chern-Simons Models},''
  arXiv:0812.3367 [hep-th].
%%%%%%%%%%%%%%%%%%%%%%%%%%%%%%%%%%%%%%%%%%%%
%%% (super) Schrodinger algebra and  rep.%%%
%%%%%%%%%%%%%%%%%%%%%%%%%%%%%%%%%%%%%%%%%%%%



%\bibitem{Perroud:1977qh}
%  M.~Perroud,
%  ``{\it Projective Representations Of The Schrodinger Group},''
%  Helv.\ Phys.\ Acta {\bf 50}, 233 (1977).

%\bibitem{Barut:1981mt}
%  A.~O.~Barut and B.~W.~Xu,
%  ``{\it Conformal covariance and the probability interpretation
%  of wave equations},'' Phys.\ Lett.\  A {\bf 82}, 218 (1981).
%
%\bibitem{Hussin:1986cc}
%  V.~Hussin and M.~Jacques,
%  ``{\it On nonrelativistic conformal symmetries and invariant tensor fields},''
%  J.\ Phys.\ A  {\bf 19}, 3471 (1986).

%\bibitem{Nishida:2007pj}
%  Y.~Nishida and D.T.~Son,
%  ``{\it Nonrelativistic conformal field theories},''
%  Phys.\ Rev.\  D {\bf 76}, 086004 (2007)
%  [arXiv:0706.3746 [hep-th]].


\bibitem{Henkel:1993sg}
  M.~Henkel,
  ``{\it Schrodinger invariance in strongly anisotropic critical systems},''
  J.\ Statist.\ Phys.\  {\bf 75}, 1023 (1994)
  [arXiv:hep-th/9310081].
%


\bibitem{Bagchi:2009my}
  A.~Bagchi and R.~Gopakumar,
  ``{\it Galilean conformal algebras and AdS/CFT},''
  arXiv:0902.1385 [hep-th].


%\cite{de Alfaro:1976je}
\bibitem{deAlfaro:1976je}
  V.~de Alfaro, S.~Fubini and G.~Furlan,
   ``{\it Conformal Invariance In Quantum Mechanics},''
  Nuovo Cim.\  A {\bf 34} (1976) 569.
  %%CITATION = NUCIA,A34,569;%%


\bibitem{Beisert:2007pu}
  N. Beisert, ``{\it
  The analytic Bethe ansatz for a chain with centrally extended $su(2|2)$
  Symmetry}",
  J.\ Stat.\ Mech. {\bf 07}, P01017 (2007),
  [arXiv:nlin.SI/0610017].




%\bibitem{Mehen:1999nd}
%  T.~Mehen, I.W.~Stewart and M.B.~Wise,
%  ``{\it Conformal invariance for non-relativistic field theory},''
%  Phys.\ Lett.\  B {\bf 474}, 145 (2000)
%  [arXiv:hep-th/9910025].


%\bibitem{Lin:2004nb}
%  H.~Lin, O.~Lunin and J.M.~Maldacena,
%  ``{\it Bubbling AdS space and 1/2 BPS geometries},''
%  JHEP {\bf 0410}, 025 (2004)
%  [arXiv:hep-th/0409174].



%%%%%%%%%%%%%%%%%%%%%%%%%%%%
%%% Superconformal Index %%%
%%%%%%%%%%%%%%%%%%%%%%%%%%%%

\bibitem{Kinney:2005ej}
  J.~Kinney, J.M.~Maldacena, S.~Minwalla and S.~Raju,
  ``{\it An index for 4 dimensional super conformal theories},''
  Commun.\ Math.\ Phys.\  {\bf 275}, 209 (2007)
  [arXiv:hep-th/0510251].

%\bibitem{Bhattacharya:2008zy}
%  J.~Bhattacharya, S.~Bhattacharyya, S.~Minwalla and S.~Raju,
%  ``{\it Indices for superconformal field theories in 3,5 and 6 dimensions},''
%  JHEP {\bf 0802}, 064 (2008)
%  [arXiv:0801.1435 [hep-th]].

\bibitem{Bhattacharya:2008bja}
  J.~Bhattacharya and S.~Minwalla,
  ``{\it Superconformal indices for ${\cal N}=6$ Chern Simons theories},''
  JHEP {\bf 0901}, 014 (2009)
  [arXiv:0806.3251 [hep-th]].

%\cite{Dolan:2008vc}
\bibitem{Dolan:2008vc}
  F.~A.~Dolan,
  ``On Superconformal Characters and Partition Functions in Three Dimensions,''
  arXiv:0811.2740 [hep-th].
  %%CITATION = ARXIV:0811.2740;%%


\bibitem{Choi:2008za}
  J.~Choi, S.~Lee and J.~Song,
  ``{\it Superconformal indices for orbifold Chern-Simons theories},''
  arXiv:0811.2855 [hep-th].




%%%%%%%%%%%%%%%%%%%%%%%%%%%%%%%%%%%%%
%%% Recent works on NR holography %%%
%%%%%%%%%%%%%%%%%%%%%%%%%%%%%%%%%%%%%

\bibitem{Nakayama:2009cz}
  Y.~Nakayama, M.~Sakaguchi and K.~Yoshida,
  ``{\it Non-Relativistic M2-brane Gauge Theory and New Superconformal Algebra},''
  arXiv:0902.2204 [hep-th].


\end{thebibliography}
 \end{document}